\pdfoutput=1
\documentclass[11pt]{article}
\usepackage[utf8]{inputenc}
\usepackage[T1]{fontenc}
\usepackage{cite}
\usepackage{hyperref}
\hypersetup{colorlinks=true,linkcolor=blue2,citecolor=red2,urlcolor=green2,pdfencoding=auto,linktocpage}
\usepackage{amsmath,amssymb,amsbsy,amstext,amsthm,simplewick,amsfonts}
\usepackage{mathrsfs}
\usepackage{graphicx}
\usepackage{wrapfig}
\usepackage{upgreek}
\usepackage{bm} 
\usepackage{framed}
\usepackage{bbm}
\usepackage{textcomp}
\usepackage{adjustbox}
\usepackage{makecell}
\usepackage{tcolorbox}
\usepackage{empheq}
\usepackage[normalem]{ulem}
\usepackage{enumitem}
\usepackage{braket} 
\usepackage{array}
\usepackage{dsfont}
\usepackage{physics}
\usepackage{ulem}
\usepackage{xcolor}
\usepackage{caption}
\usepackage{subcaption}
\usepackage{tocloft}
\usepackage{tikz}
\usetikzlibrary{decorations.pathmorphing,decorations.markings}
\usetikzlibrary{arrows.meta}
\usetikzlibrary{positioning}
\usepackage{circuitikz}
\usepackage{booktabs}
\usepackage{pifont}

\usepackage[framemethod=default]{mdframed}

\newmdenv[backgroundcolor=gray!15,%
skipabove=5pt,%
skipbelow=5pt,%
leftmargin=2pt,%
rightmargin=2pt,%
innertopmargin=-6pt,%
innerbottommargin=5pt,%
innerleftmargin=5pt,%
innerrightmargin=5pt,%
splittopskip=0pt,%
splitbottomskip=0pt,%
linewidth=0pt,%
nobreak=true]%
{keyeqn}

\newmdenv[backgroundcolor=gray!15,%
skipabove=5pt,%
skipbelow=5pt,%
leftmargin=2pt,%
rightmargin=2pt,%
innertopmargin=5pt,%
innerbottommargin=5pt,%
innerleftmargin=5pt,%
innerrightmargin=5pt,%
splittopskip=0pt,%
splitbottomskip=0pt,%
linewidth=0pt,%
nobreak=true]%
{keyeqn2}

\graphicspath{{Fig/}}

\definecolor{red2}{RGB}{214, 39, 40}
\definecolor{green2}{RGB}{0,170,0}
\definecolor{blue2}{RGB}{0,100,200}
\definecolor{magenta2}{RGB}{191,64,191}
\definecolor{purple2}{RGB}{112,48,160}
\definecolor{orange2}{RGB}{255,192,0}

\newcommand{\dof}{\#_\text{d.o.f.}}

\newcommand{\G}{\mathcal{G}}
\newcommand{\E}{\mathcal{E}}
\newcommand{\Nr}{N_r}
\newcommand{\Na}{N_a}

\newcommand{\C}{\mathcal{C}}
\newcommand{\SSK}{S_\textrm{MSR}}
\newcommand{\then}{\qquad \Rightarrow \qquad}
\newcommand{\pa}{\phi^a_i}
\newcommand{\pr}{\phi^r_I}
\newcommand{\ex}[1]{\langle #1 \rangle}

\numberwithin{equation}{section}

\allowdisplaybreaks[1]
\begin{document}

\begin{titlepage}
	\setcounter{page}{1} \baselineskip=15.5pt 
	\thispagestyle{empty}

     \begin{center}
		{\fontsize{18}{18}\centering {\bf Counting Degrees of Freedom\\[0.2cm] in Open Effective Theories}} 
	\end{center}
 
	\vskip 18pt
	\begin{center}
		\noindent
		{\fontsize{12}{18}\selectfont Enrica Lausdei\footnote{\tt enrica.lausdei2001@gmail.com}$^{,a,b}$, and Enrico Pajer\footnote{\tt enrico.pajer@gmail.com}$^{,c}$}
	\end{center}
	
	\begin{center}
        \vskip 12pt
		$^{a}$ \textit{Fakult\"at f\"ur Physik, Ludwig-Maximilians-Universit\"at M\"unchen,\\
Theresienstr.~37, 80333 M\"unchen, Germany\\}           \vskip 12pt
		$^{b}$ \textit{Physik Department T31,
James-Franck-Stra{\ss}e 1,\\ Technische Universit\"at M\"unchen,
D--85748 Garching, Germany\\}
		\vskip 12pt
		$^{c}$ \textit{Department of Applied Mathematics and Theoretical Physics, University of Cambridge,\\Wilberforce Road, Cambridge, CB3 0WA, UK} 
	\end{center}

	    \vskip 40pt
		
		\noindent\rule{\textwidth}{0.5pt}
		\noindent \textbf{Abstract} ~~      

Open effective field theories provide a systematic framework for describing physical systems interacting with an environment whose microscopic details are unknown, unobservable, or uncalculable. A basic step in constructing any effective field theory is the identification of the relevant degrees of freedom. For open effective theories, however, this step is subtle: their dynamics is generically dissipative and non-Hamiltonian, so the standard Hamiltonian and Lagrangian algorithms are not directly applicable.

To overcome this limitation, we develop an algorithm to count degrees of freedom directly from the equations of motion. Restricting to classical, linearised dynamics on homogeneous and isotropic backgrounds, our method applies to non-Lagrangian systems with unequal numbers of fields and equations, including systems with constraints and gauge redundancies. As a by-product, our algorithm identifies constraints, gauge identities, gauge redundancies, and consistency conditions on stochastic sources. The central ingredient is the introduction of a dual set of ``advanced'' equations, or equivalently an auxiliary Martin-Siggia-Rose functional. We show that gauge redundancies of the original fields are associated with gauge identities of the dual advanced equations, and vice versa. We illustrate our procedure in examples ranging from coupled scalar systems to electromagnetism in a medium and gravitational effective theories relevant for cosmology. Our results will prove useful in the study of stochastic dynamics, non-equilibrium statistical systems and the semi-classical limit of open quantum systems on the Schwinger-Keldysh path integral.

	
\end{titlepage} 


\newpage
\setcounter{page}{2}
{
	\tableofcontents
}

\section{Introduction}\label{IntroSect}

Effective field theories provide one of the most powerful organising principles in physics. Their usefulness rests on a simple but far-reaching observation: when a physical problem exhibits a separation of scales, the low-energy or long-distance phenomena can often be described without keeping track of all microscopic details. Instead, one writes the most general dynamics for the relevant low-energy degrees of freedom, organised in an expansion in derivatives, fields, or inverse powers of the heavy scale. This idea has led to remarkable simplifications in a wide range of areas, from condensed matter physics to particle physics, gravity and cosmology.

The starting point in constructing any effective theory is the identification of the degrees of freedom that the theory is meant to describe. By number of degrees of freedom we mean half of the initial conditions of the equations of motion that are needed to fix the physical evolution of the system. This is especially important and subtle in systems with constraints or gauge redundancies, where not every component of the fields corresponds to an independent physical degree of freedom. For unitary conservative theories, which are the main focus of much of the high-energy physics literature, there are several well-established procedures to address this question. Perhaps the most famous is the Dirac-Bergmann algorithm \cite{Dirac:1950pj,Anderson:1951ta,Wipf:1993xg} in the Hamiltonian formalism, which gives a systematic way to identify constraints, distinguish primary from secondary constraints, classify them into first and second class, and ultimately count the number of physical degrees of freedom.  This analysis also provides the starting point for canonical quantisation. On the Lagrangian side, related algorithms identify constraints, gauge identities and gauge transformations directly from the equations of motion, and have been used extensively in the study of gauge theories and higher-derivative systems \cite{Henneaux:1992ig,Rothe:2010dzf_dof}.

The situation is less straightforward for open systems. By an open system we mean a system whose effective dynamics is obtained after unobserved or unknown environmental degrees of freedom have been integrated out. The resulting equations of motion need not be Hamiltonian and may contain dissipative terms. However, the very feature that makes open systems interesting also makes the standard counting of degrees of freedom subtle. Since dissipative equations do not in general arise from an ordinary Hamiltonian, the Dirac--Bergmann algorithm cannot be applied directly. Indeed, for example, the absence of a symplectic structure means the phase space need not be even dimensional. Similarly, the Lagrangian formula does not correctly count the physical degrees of freedom once the equations are not the Euler--Lagrange equations of a unitary action. Given the continuously increasing interest in open effective field theories \cite{LopezNacir:2011kk,Crossley:2015evo,Glorioso:2017fpd,Boyanovsky:2018fxl,Boyanovsky:2015tba,Boyanovsky:2015jen,Liu:2018kfw,Hongo:2018ant,Hongo:2019qhi,Hongo:2024brb,Liu:2024tqe,Burgess:2024heo,Colas:2024lse ,Salcedo:2024nex,Lau:2024mqm,Cespedes:2025ple,Burgess:2014eoa,Burgess:2015ajz,Burgess:2022rdo, Vardhan:2024qdi,Zarei:2021dpb,Ota:2024mps,Mirbabayi:2024eml,Green:2025hmo,Sharifian:2023jem,Kaplanek:2025moq,Colas:2025ind,Lopez:2025arw,Salcedo:2026cqb,Kaplanek:2026kpp}, it is important to overcome this limitation. 

The goal of this paper is to develop a simple algorithm for counting degrees of freedom in open systems. We do not attempt to treat the most general system or quantum dynamics. Instead, we restrict our attention to a more modest, but still broadly useful, class of systems. First, we consider \textit{classical} deterministic dynamics. Second, we focus on \textit{linear} equations of motion, namely free theories. Third, although our intended applications are field theories, we formulate the algorithm for \textit{ordinary} differential equations in time, rather than partial differential equations. Thus, the mathematical problem we address is the following: given a set of $N_a$ linear equations
\begin{align}
    E_i(\phi)=0\,,
    \qquad
    i=1,\dots,\Na\,,
\end{align}
for a set of $\Nr$ fields $\phi_I$, with $I=1,\dots,\Nr$, how many physical degrees of freedom do these equations propagate?

At first sight the restriction to classical, linear odes may seem severe\footnote{Recent work has also revisited the problem of counting degrees of freedom directly at the level of the field equations. In particular, in \cite{Heisenberg:2025fxc} L.~Heisenberg develops a more general method based on the Cartan--Kuranishi analysis of differential equations, building on earlier ideas of Einstein, Hilbert, Cartan, Kuranishi and Seiler. That approach is neither Hamiltonian nor Lagrangian, and applies directly to the non-linear field equations. A direct comparison to our results would be interesting.}, but it is less limiting than it appears. Effective field theories are usually organised perturbatively around a free theory. The free quadratic part determines the spectrum, the number of propagating degrees of freedom and the power counting used to organise the interacting theory. Therefore, even when the full effective theory is nonlinear, the correct counting of degrees of freedom is already fixed at the level of the linearised equations. Moreover, in many applications the background is homogeneous and isotropic. After Fourier transforming in space, local spatial derivatives become multiplication by powers of the wave number, and the linearised partial differential equations reduce to ordinary differential equations in time. Finally, the restriction to classical deterministic dynamics is natural in semiclassical applications of open quantum systems, where the dominant physics is captured by a saddle-point evolution. Indeed, as we discuss in Section \ref{sec2p3}, the Martin-Siggia-Rose (MSR) functional \cite{PhysRevA.8.423,Janssen:1976qag,DeDominicis:1976} appearing in our construction often emerges in the semi-classical limit of the Schwinger-Keldysh path integral \cite{Schwinger:1960qe,Keldysh:1964ud}. 

Our construction starts from re-writing and re-labelling the above the equations of motion as
\begin{align}
    E_i^r=M_{iI}\phi_I^r=0\,,
\end{align}
where the matrix of differential operators $M_{iI}$ need not be square, since $N_r$ and $N_a$ are allowed to be different. For later convenience, we borrow nomenclature from the MSR formalism and label the equations $E_i=E_i^r$ and the fields appearing in them, $\phi_I=\phi_I^r$, with and $r$ standing for ``retarded''. We first apply a Lagrangian-style constraint algorithm directly to these equations. This identifies the number of constraints on the retarded initial data, which we denote by $l_r$. It also detects possible retarded gauge identities, namely linear combinations of the equations of motion and their derivatives that vanish identically, i.e.~off the mass shell. The crucial new step is to associate to the retarded equations a dual set of equations, which we dub ``advanced'' equations. Equivalently, we introduce an auxiliary MSR action 
\begin{align}
    \SSK=\int \phi_i^a E_i^r
    =\int \phi_i^a M_{iI}\phi_I^r\,,
\end{align}
for new ``advanced'' fields $\pa$ and define the advanced equations by varying with respect to the retarded fields,
\begin{align}
    E_I^a=(M^\dagger)_{Ii}\phi_i^a=0\,.
\end{align}
Note that by construction we have $N_a$ advanced fields and $N_r$ advanced equations. Running the same  algorithm on the advanced equations gives a number $l^a$ of advanced constraint equations, a number $g^a$ of advanced gauge identities and a corresponding number $e_a$ of gauge transformations.

The main result of this paper is that the number of degrees of freedom for the original set of equations is
\begin{align}\label{mainresult}
    \dof
    =
    \Nr-\frac12\left(l_r+g_a+e_a\right)\,.
\end{align}
This differs from the standard unitary formula \eqref{dofunitary}
in an important way. As expected, the constraints that restrict the allowed retarded initial data are the retarded constraints and therefore contribute through $l_r$. However, the gauge transformations acting on the retarded fields are generated not by the $g_r$ retarded gauge identities, but by the $g_a$ advanced gauge identities. This follows directly from the MSR action: after integration by parts, a transformation of the retarded fields is related to an identity among the advanced equations of motion. Thus the relevant gauge data for the retarded fields are $g_a$ and $e_a$, rather than $g_r$ and $e_r$.

In the body of the paper, we illustrate the algorithm in a sequence of examples. We begin with simple scalar theories, where the counting can be checked explicitly. We then turn to electromagnetism, both before and after gauge fixing, using it as a benchmark for the treatment of constraints and gauge redundancies. We also discuss systems where a naive application of the unitary counting formula gives the wrong answer, while our retarded/advanced prescription gives the expected result. Finally, we apply the method to gravitational effective theories. In general, the distinction between retarded and advanced gauge identities is important for obtaining the correct number of degrees of freedom.\\

The rest of the paper is organised as follows. In Section \ref{sec:2p1} we review the constraint algorithm that extracts constraints, gauge identities and gauge transformations from a set of linear equations of motion. We then explain how this algorithm is modified for open systems by introducing the dual advanced equations. We clarify that our algorithm applies to generic classical equations, possibly with stochastic sources. This large class of theories includes the semi-classical limit of open quantum systems, as made clear in the Schwinger-Keldysh formulation. In Section \ref{sec:3}, we elaborate on the role of the advanced gauge symmetry in constraining the stochastic sources that may be added to a set of deterministic equations and in reducing those equations to a minimal subset. In Section \ref{sec:ex}, we apply our procedure to progressively more involved examples, culminating in gravitational effective theories relevant for cosmology. We conclude with a discussion of possible extensions of our work.

As an invitation for the reader, we briefly discuss now a simple toy model that displays the general elements of our algorithm. 


\subsection{A toy model}

To motivate and demonstrate our analysis, we study a simple toy model of two coupled scalars. This toy model is simple enough that the final result can be easily guessed and, at the same time, sufficiently complex that it displays many of the general features we will discuss in the rest of this work. 

Let's proceed in steps. First, consider the following two ordinary differential equation
\begin{align} 
\ddot \phi +\lambda \phi &=0 \,, & \ddot \chi + \chi &=0\,,
\end{align}
for two variables $\phi(t)$ and $\chi(t)$ with $\lambda$ some real constant. This theory requires four initial conditions, namely $ (\phi(t_0),\dot \phi(t_0),\chi(t_0),\dot \chi(t_0))$ and therefore we say it has $4/2=2$ degrees of freedom. Now we make things interesting by adding a third equation:
\begin{align}
    E^r_1 &\coloneqq \ddot \phi+\lambda \phi=0\,,
    &
    E^r_2 &\coloneqq \ddot \chi+\chi=0\,,
    &
    E^r_3 &\coloneqq \dot \phi+\chi=0\,.
\end{align}
How many degrees of freedom are there now? A brute force way to answer this question would be to solve the system and count the number of initial conditions. Instead, we will approach this by mimicking the procedure that then extends to the most general case. 

First we notice that, for generic values of $\lambda$, this system is overconstrained. Indeed, using $E^r_3=0$ we have $\chi=-\dot\phi$, and therefore $E^r_2=0$ gives
\begin{align}
    \dddot\phi+\dot\phi=0\,.
\end{align}
On the other hand, differentiating $E^r_1=0$ gives
\begin{align}
    \dddot\phi+\lambda \dot\phi=0\,.
\end{align}
Thus, unless $\lambda$ is tuned to one, the equations force the solution to be trivial. The insightful reader could have guessed this from the start: if the two harmonic oscillators $\phi$ and $\chi$ do not oscillate with the same frequency, they cannot continue satisfying the constraint $\dot \phi+\chi=0$ at all times. 

The non-trivial case relevant for us is $ \lambda=1$ and in the following we restrict to this value. Let us first analyse the ``retarded'' equations and define
\begin{align}
    \phi^r_I=(\phi,\chi)\,,
    \qquad I=1,2\,.
\end{align}
At the first step we note that the third equation is a constraint equation. Equivalently, we can say that for $v_i=(0,0,1)$ one finds the constraint
\begin{align}
    \C_1 \coloneqq v_i E^r_i= E^r_3=\dot\phi+\chi=0\,.
\end{align}
This is not the end of the analysis because we have to ensure that the constraint is satisfied at all times. Hence, we add the time derivative of this constraint to the system, which now is $(E^r_i,\dot \C_1)$. Combining $\dot \C_1$ with $E^r_1$ gives a second constraint,
\begin{align}
    \C_2
    \coloneqq (-1,0,0,1)\cdot (E^r_i,\dot \C_1)=
    \dot \C_1-E^r_1
    =
    \dot\chi-\phi=0\,.
\end{align}
We now iterate by adding the time derivative of $\C_2$ to our system of equations, which now becomes $(E^r_i,\dot \C_1,\dot \C_2)$ with
\begin{align}
    \dot \C_2=\ddot\chi-\dot\phi\,.
\end{align}
We now notice that we can find linear combinations of these equations that do not involve second time derivatives, as we did before, and therefore should be interpreted as constraints: 
\begin{align}
    \dot \C_2-E^r_2=\ddot \chi -\dot \phi -(\ddot \chi +\chi)=-(\dot \phi+\chi)=0ù,.
\end{align}
However, this constraint is a linear combinations of the previous ones
\begin{align}
    \dot \C_2-E^r_2+E^r_3
    =
    \ddot\chi-\dot\phi
    -(\ddot\chi+\chi)
    +(\dot\phi+\chi)
    \equiv 0\,.
\end{align}
Equivalently, written directly in terms of the original equations, this is the identity
\begin{align}
    \G_r
    \coloneqq
    \ddot E^r_3-\dot E^r_1-E^r_2+E^r_3
    \equiv 0\,.
    \label{eq:toy-retarded-gauge-identity}
\end{align}
This is an \textit{off-shell} identity in the sense that it is valid irrespectively of whether the equations of motion are satisfied. We summarise this analysis into three numbers: the number of constraints $l_r$, the number of gauge identities $g_r$, and the number $e_r$ equal to one plus the largest number of time derivatives appearing in the gauge identity. Our analysis gives
\begin{align}
    l_r=2\,,
    \qquad
    g_r=1\,,
    \qquad
    e_r=3\,.
\end{align}
If we substituted this into the standard formula for the counting of degrees of freedom for unitary theory, \eqref{dofunitary}, we would find a negative number of degrees of freedom, which is incorrect even when interpreted as a zero. Instead we note the following. 

When studying Euler-Lagrange equations from a Lagrangian, for which one always has $N_r=N_a$, gauge identities are related to gauge transformations acting on the fields. For open systems, there is an interesting twist. The identity \eqref{eq:toy-retarded-gauge-identity} does \textit{not} imply a gauge redundancy of the retarded fields. Instead, it implies a gauge redundancy of the advanced fields. To make this precise, introduce an auxiliary MSR-like action,
\begin{align}
    \SSK
    =
    \int dt\,
    \left[
        \psi_1 E^r_1
        +
        \psi_2 E^r_2
        +
        \psi_3 E^r_3
    \right]\,,
    \label{eq:toy-sk-action}
\end{align}
where the $\psi_i$ are advanced fields. Explicitly,
\begin{align}
    \SSK
    =
    \int dt\,
    \left[
        \psi_1(\ddot\phi+\phi)
        +
        \psi_2(\ddot\chi+\chi)
        +
        \psi_3(\dot\phi+\chi)
    \right]\,.
\end{align}
Up to boundary terms this can be written as
\begin{align}
    \SSK
    \simeq
    \int dt\,
    \left[
        -\dot\psi_1\dot\phi+\psi_1\phi
        -\dot\psi_2\dot\chi+\psi_2\chi
        +
        \psi_3(\dot\phi+\chi)
    \right]\,.
\end{align}
The retarded gauge identity \eqref{eq:toy-retarded-gauge-identity} implies the following gauge redundancy of the advanced fields:
\begin{align}
    \delta\psi_1 &= \dot\epsilon\,,
    &
    \delta\psi_2 &= -\epsilon\,,
    &
    \delta\psi_3 &= \ddot\epsilon+\epsilon\,.
    \label{eq:toy-advanced-gauge-transformation}
\end{align}
The MSR functional is indeed invariant under this transformation for any $\epsilon(t)$,
\begin{align}
    \delta \SSK
    &=
    \int dt\,
    \left[
        \dot\epsilon\,E^r_1
        -
        \epsilon\,E^r_2
        +
        (\ddot\epsilon+\epsilon)E^r_3
    \right]
    \nonumber\\
    &\simeq
    \int dt\,\epsilon
    \left[
        \ddot E^r_3-\dot E^r_1-E^r_2+E^r_3
    \right]
    =0\,,
\end{align}
where again we dropped boundary terms. Therefore the retarded gauge identity gives a gauge redundancy of the auxiliary advanced fields, not of the physical retarded fields. As we see now, this implication is symmetric under the exchange of advanced and retarded. 

We now derive the advanced equations by varying \eqref{eq:toy-sk-action} with respect to the retarded fields. Varying with respect to $\phi$ and $\chi$ gives
\begin{align}
    E^a_\phi
    &\coloneqq
    \frac{\delta \SSK}{\delta \phi}
    =
    \ddot\psi_1+\psi_1-\dot\psi_3=0\,,
    \\
    E^a_\chi
    &\coloneqq
    \frac{\delta \SSK}{\delta \chi}
    =
    \ddot\psi_2+\psi_2+\psi_3=0\,.
    \label{eq:toy-advanced-eoms}
\end{align}
It is manifestly impossible to combine these two equations to remove the second time derivatives to obtain a constraint. 
Hence the advanced algorithm terminates immediately and gives
\begin{align}
    l_a=0\,,
    \qquad
    g_a=0\,,
    \qquad
    e_a=0\,.
\end{align}
According to \eqref{mainresult}, the number of retarded degrees of freedom is therefore
\begin{align}
    \dof
    =
    \Nr-\frac12(l_r+g_a+e_a)
    =
    2-\frac12(2+0+0)
    =
    1\,.
\end{align}
This agrees with the explicit solution of the system. The equation $E^r_3=0$ gives
\begin{align}
    \chi=-\dot\phi\,,
\end{align}
and then $E^r_1=0$ gives
\begin{align}
    \ddot\phi+\phi=0\,.
\end{align}
The field $\chi$ is therefore fixed in terms of $\phi$, and the system propagates precisely one harmonic-oscillator degree of freedom. This example illustrates that, for a non-Lagrangian set of equations, the retarded gauge identities do not imply gauge redundancies of the retarded fields.

There is another way to view the same result. The identity
\begin{align}
    \ddot E^r_3-\dot E^r_1-E^r_2+E^r_3\equiv0ù,,
\end{align}
contains $E^r_2$ without time derivatives. Therefore $E^r_2$ can be solved for algebraically in terms of the other equations and their derivatives:
\begin{align}
    E^r_2
    =
    \ddot E^r_3-\dot E^r_1+E^r_3\,.
\end{align}
Thus the equation $E^r_2=0$ is redundant once $E^r_1=0$ and $E^r_3=0$ are imposed. The system may equivalently be written in the minimal form
\begin{align}\label{minimal}
    \ddot\phi+\phi=0\,,
    \qquad
    \dot\phi+\chi=0\,,
\end{align}
which makes the single degree of freedom manifest.

Finally, the advanced gauge symmetry also constrains stochastic generalisations of these differential equations, as for example in the Langevin equation. Indeed, suppose the deterministic equations in this model are promoted to sourced equations
\begin{align}
    E^r_i=J_i\,.
\end{align}
Here $J_i$ may include fixed, deterministic external sources or stochastic external sources with some assigned probability distribution. The retarded gauge identity implies the corresponding ``noise'' constraint
\begin{align}
    \ddot J_3-\dot J_1-J_2+J_3=0\,.
    \label{eq:toy-noise-constraint}
\end{align}
Thus the advanced gauge symmetry does not reduce the number of retarded degrees of freedom; rather, it restricts the allowed form of sources and stochastic noise. In this example, the advanced gauge symmetry also instructs us on how to reduce the original system of equations to an equivalent minimal subset as in \eqref{minimal}.


\paragraph{Notation and conventions} We adopt the mostly plus signature, $(-,+,+,+)$ and for converting derivatives from position to Fourier space, we use 
\begin{align}
    \partial_\mu \to (-i\omega,i\mathbf{k})\,.
\end{align}
We often Fourier transform only the spatial coordinates, but not the time coordinates. We use the symbol ``$\coloneqq$'' for definition and the symbol ``$\equiv$'' for equalities that are valid off-shell, i.e.~(gauge) identities. \\


\section{The algorithm}

Consider a set of $\Nr$ fields, $\phi_I$ with $I=1,\dots,\Nr$. These fields obey a set of $\Na$ \textit{linear} second-order ordinary differential equations in time, which we denote by
\begin{align}\label{eom}
    E_i(\phi)=0\,, \quad \text{with} \quad i =1,\dots , N_a\,.
\end{align}
The labels $r$ and $a$ are only notation: $N_r$ denotes the number of fields, while $N_a$ denotes the number of equations. They are inspired by the terminology of the MSR formalism. No prior knowledge of the MSR formalism and of the associated Schwinger-Keldysh path integral will be needed to understand our results. 

In many cases of interest, the equations \eqref{eom} do not uniquely determine all components of the fields. More precisely, there may exist transformations of the fields,
\begin{align}
    \phi_I(t) \mapsto \phi_I(t) + \delta \phi_I(t)\,,
\end{align}
which map any solution of \eqref{eom} to another solution of the same equations, without changing the corresponding physical state. We shall call such transformations gauge transformations. The physical solution space is then defined as the quotient of the space of solutions by these gauge transformations: two solutions are identified if they are related by a gauge transformation.

We would like to determine how many independent initial conditions are required to specify a physical solution. For a linear system, or for the linearisation of a nonlinear system around a fixed background, the space of solutions is a vector space, and the physical solution space is the quotient vector space obtained after modding out by gauge transformations. The number of independent initial conditions is the dimension of this physical solution space.


We define the \textit{number of degrees of freedom} $\dof$ to be half of the number of independent initial conditions:
\begin{align}
   \dof = \frac12 \text{ Number of initial conditions}\,.
\end{align}
When the equations above come from the variation of an action that is given as an integral of a local Lagrangian, there is a well-known algorithm that counts the number of degrees of freedom \cite{Rothe:2010dzf_dof}. 
A necessary condition for this to be possible is that $\Na=\Nr$, since the variation of the action with respect to each one of the fields gives one additional equation. The formula for the number of degrees of freedom in this case requires the calculation of three independent integers:
\begin{itemize}
    \item $l$: This is the \textit{number of constraints} on the possible initial conditions. Intuitively, this is related to the number of independent equations that can be written that do not involve second time derivatives of the fields. 
    \item $g$: This is the \textit{number of gauge identities}, namely linear combinations of the equations of motion and their time derivatives that vanish identically, irrespectively of whether the equations of motion are satisfied or not. Intuitively, a non-vanishing $g$ indicates that there are actually fewer equations of motion than $\Nr$.
    \item $e$: This is the \textit{number of gauge transformations} that leave the equations invariant. Gauge transformations involving up to $n$ time derivatives of the gauge parameter contribute an amount $n$ to the integer $e$. 
\end{itemize}
The formula for the number of degrees of freedom for unitary theories in terms of these three numbers is simply
\begin{align}\label{dofunitary}
    \text{Unitary: } \quad \dof=N_r-\frac12(l+g+e)\,.
\end{align}
When the equations of motion do not derive from the variation of an action (which is always the case, in particular when $N_r\neq N_a$), then this formula does not give the correct result. We will explain how to find the correct result shortly. Nevertheless, the calculation of $l$, $g$ and $e$ from a given set of equations $E_i=0$ is still an important step in the procedure, so we explain it next.


\subsection{Finding $l$, $g$ and $e$}\label{sec:2p1}

In the following, we outline an algorithm that, starting from a set of $N_a$ equations of motion $E_i=0$ as in \eqref{eom}, outputs a set of three integers $l$, $g$ and $e$. To begin with, we rewrite these equations as follows: 
\begin{align}
    E_i=W_{iJ} \ddot{\phi}_J + K_i(\phi,\dot \phi)=0\,,
\end{align}
where $W$ is a rectangular matrix of size $N_a \times N_r$ and $K_i$ are $N_a$ linear functions of all the fields and their first derivatives. 

\paragraph{Step 0} Now we define the zeroth step of the algorithm. As we will see shortly, the subsequent steps will be iterations of this step, starting from an enlarged set of equations of motion. Each step will consist of the following operations:
\begin{itemize}
    \item Find the left null vectors of $W$ and hence a set of constraints $\C$.
    \item Determine if the constraints are linearly independent; if they are not, find the combination of them that vanishes identically, which are called gauge identities $\G$. 
    \item From each gauge identity, derive a corresponding gauge transformation. 
\end{itemize}
For Step 0 we start by finding the left e-vectors vector of $W$ with e-value zero, i.e.~satisfying
\begin{align}
    v_i W_{iJ}=0\,.
\end{align}
If no such left e-vector exists, then the algorithm stops. 
For each of these e-vectors $v_i$, we obtain an equation by contracting it with the equations of motion:
\begin{align}
    \C \coloneqq v_i E_i = v_i K_i=0\,.
\end{align}
All such equations are \textit{constraints} in the sense that they depend only on the fields and their first derivatives, but not on their second derivatives. These represent restrictions on the set of initial conditions that are consistent for the system of equations. In general, we can have many different $\C$'s corresponding to the many different $v$'s. We choose not to make explicit the additional label that counts each of these constraints to keep our notation as simple as possible, but the reader should remember that in general $\C$ is a set of constraints. These constraints could be linearly independent or not. If they are not, then there must be linear combinations such that 
\begin{align}
    \G\coloneqq \sum_\alpha a_\alpha \C_\alpha \equiv 0 \quad \text{(off-shell)}\,.
\end{align}
Such a relation is called a \textit{gauge identity}. It is an identity in the sense that it holds also when the equations of motion are not satisfied. It is an off-shell relation. Again, there could be many gauge identities that are linearly independent of each other, but we omit the label that would count each of them to keep the notation simple. Each linearly-independent gauge identity contributes 1 to the integer $g$. If all the constraints corresponding to the set of left null vectors are related to previous constraints by gauge identities, the algorithm stops. 
The relation between gauge identities from the gauge transformation will be discussed shortly, but first we would like to explain how to go from one step to the next. 


\paragraph{Step $n$} The operations described above in Step 0 are now repeated on an enlarged set of equations of motion defined as follows. Let's assume that we have executed Step $n-1$ and the algorithm has not stopped. This in particular means that Step $n-1$ finished with at least some new linearly-independent constraints. In this case, we define the starting point of Step $n$ to be all the equations of motion $E_i^{(n-1)}=0$ of the previous step plus all the time derivatives $\dot \C^{(n-1)}=0$ of the new constraints that we found:
\begin{align}
    E_{j}^{(n)}=\{E_{i}^{(n-1)},\dot \C^{n-1}\} =0\,.
\end{align}
Now we simply apply the three operations above to the input $E^{(n)}_j$ of Step $n$. This will yield a new set of left null vectors $v^{(n)}_j$, a corresponding set of constraints $\C^{(n)}=0$, and possibly a set of gauge identities $\G^{(n)}=0$. 

Now we notice that gauge identities are linear combinations of the constraints, which in turn are linear combinations of the original equations of motion and their time derivatives. Therefore, any gauge identity can be written in the following form:
\begin{align}\label{gaugeid}
    \G=\sum_{i=1}^{N_a} \sum_{m=0}^M\rho_{i,m}\left( \frac{d}{dt} \right)^m E_i \equiv 0\,,
\end{align}
where, again, this equation is valid off-shell. Here, the $\rho_{i,m}$ are coefficients, and the sum over $m$ will go up to some finite maximum number $M$ of derivatives. Each such gauge identity contributes $(M+1)$ to the integer $e$ and 1 to the integer $g$,
\begin{align}
    \G=\sum_{i=1}^{N_a} \sum_{m=0}^M\rho_{i,m}\left( \frac{d}{dt} \right)^m E_i \equiv 0 \then g\to g+1\,, \quad e\to e+1+M\,.
\end{align}
Note that, by construction, the following inequality holds:
\begin{align}
    e\geq g\,.
\end{align}
We will see shortly that $e$ is counting the number of gauge transformations, but let's set this aside for the moment. 

We can denote the number of new constraints, the number of new gauge identities and the number of gauge transformations of Step $n$ by $l^{(n)}$, $g^{(n)}$, and $e^{(n)}$ respectively. If at Step $n$ no new left null-vectors are found, or all of the constraints are related to previous ones by gauge identities, the algorithm stops. When the algorithm has stopped, we should add up all of the contributions to the number of constraints and the number of gauge identities from each step
\begin{align}
    l& \coloneqq \sum_n l^{(n)}\,, & g& \coloneq\sum_n g^{(n)}\,, & e& \coloneq \sum_n e^{(n)}\,.
\end{align}
It is possible that the algorithm never stops and the number of constraints $l$ and/or of gauge identities $g$ keeps increasing forever. To prevent this, we add the additional rule that, when $l$ and $g$ after $n$ steps are bigger than $2 N_r$, the algorithm must be stopped. If this happens, we conclude that the original set of equations does not admit any non-trivial solution, namely the only possible solution is the one in which all fields vanish up to gauge transformations. 


\paragraph{Gauge transformations in unitary theories}

In the algorithm described above, we found a number of gauge identities, and we said that they were associated to gauge transformations that are counted by the integer $e$. It is time to explain where this connection comes about: first we do this for unitary theories that come from a Lagrangian; then we generalise this in the next section to open theories that come from a MSR action. 

For unitary theories whose equations come from a Lagrangian, the gauge identity in \eqref{gaugeid} immediately tells us that the transformation of the fields 
\begin{align}\label{gaugetransfUni}
    \Delta \phi_i=\sum_{m=0} (-1)^m\left(\frac{d}{dt}\right)^m\left(\rho_{im} \epsilon\right)\,.
\end{align}
is a gauge transformation, namely, it leaves the action invariant up to boundary terms. To see this, we simply notice that, for quadratic action that leads to the equations of motion $E_i=0$, one must have
\begin{align}\label{unitaryaction}
    S=\int \phi_i M_{ij} \phi_j+\text{b.t.}=\int \phi_i E_i+\text{b.t.}\,,
\end{align}
where ``b.t.'' denote boundary terms that do not contribute to the equations of motion. Inserting the gauge transformation \eqref{gaugetransfUni} and integrating by parts gives back the gauge identity up to boundary terms, hence proving gauge invariance:
\begin{align}
    S\simeq \int \Delta \phi_i E_i \simeq \int \epsilon\, \G =0\,,
\end{align}
where $\simeq$ denotes equality up to boundary terms. It should be noted that this derivation is completely equivalent to the perhaps more familiar Hamiltonian analysis that involves primary, secondary, etc. constraints, as well as the classification into first and second class constraints. This Hamiltonian analysis has been discussed many times in a large number of references, such as, for example \cite{Henneaux:1992ig}. We refrain from discussing it here as our goal is to develop a formalism that can be used even when a Hamiltonian does not exist, such as for open dissipative theories.




\subsection{Dissipative theories}

We shift our focus now to general open dissipative theories. These theories do not come from an ordinary Lagrangian, and therefore the formula \eqref{dofunitary} for the counting of degrees of freedom gives the incorrect results in general. Moreover, the discussion around \eqref{unitaryaction} does not apply, and one needs a different procedure, which we now explain. 

Our starting point is again a set of $N_a$ equations of motion for $N_r$ fields. It will be useful to refer to equations as \textit{retarded equations of motion} and to the fields as \textit{retarded fields}, to distinguish them from other equations and fields that we will encounter shortly. For clarity, we will therefore henceforth introduce a label $r$, which stands for retarded. Since the equations are linear by assumption, we can write this as 
\begin{align}\label{retardedeom}
    E_i^r=M_{iI} \phi^r_I=0 \qquad \text{ for } i=1,2,\dots,N_a\,, \text{ and } I=1,2,\dots,N_r\,.
\end{align}
where $M_{iI}$ is a generically rectangular $N_a \times N_r$ matrix. Note that we do not assume any relation between the number of equations and the number of fields, and in particular, the following procedure applies also when $N_a \neq N_r$.

\paragraph{The dual equations of motion} To any given set of retarded equations of motion, we can associate a ``dual'' or ``conjugate'' set of equations of motion, which we will call \textit{advanced equations of motion} and denote by $E^a_I$, with $I=1,\dots,N_r$:
\begin{align}
    E_i^r = M_{iI}\phi^r_I=0 \quad \leftrightarrow \quad E^a_I=(M^{\dagger})_{Ii}\phi^a_i=0\,.
\end{align}
As suggested by the above notation, this duality operation is simply the Hermitian conjugation of the linear operator $M$ appearing in the linear equations of motion. As such, if we perform this duality twice, we recover the original equations. 

This construction has a natural interpretation in terms of the MSR functional and the semi-classical limit of the Schwinger-Keldysh path integral. In full generality, we can always think that the retarded equations arise from the variation of a MSR action of the form 
\begin{align}\label{SSK}
 \SSK = \int \pa E_i^r =\int \phi^a_i  \, M_{iI} \, \phi^r_I\,,
 \end{align}
where $\pa$ are auxiliary fields. We will call $\pa$ ``advanced'' fields\footnote{In the discussion of open systems, these fields arise by going to the Keldysh basis after the doubling of degrees of freedom in the closed time contour, but none of this information is necessary for the following discussion.}. The equations of motion now simply arise from varying the action with respect to the advanced fields
\begin{align}
    \frac{\delta \SSK}{\delta \pa}=E_i^r=0\,.
\end{align}
By following the algorithm described in Section \eqref{sec:2p1} from the equations $E_i^r=0 $, we can extract the three integers $l_r$, $g_r$ and $e_r$, which we also denote with an upper label $r$ standing for retarded,
\begin{align}
    E_i^r=0 \then (l_r,g_r,e_r)\,.
\end{align}
To proceed, one is instructed to now vary the MSR action with respect to the retarded fields to find what we will call the advanced equations of motion
\begin{align}
        \frac{\delta \SSK}{\delta \pr}=E_I^a=0\,.
\end{align}
This is a set of $N_r$ linear equations, which can also be written as 
\begin{align}
    (M^\dagger)_{Ii}\pa=0\,,
\end{align}
where $M^\dagger$ is the Hermitian conjugate operator of $M$ appearing in \eqref{retardedeom}. In practise, this is just obtained by integrating by part the action $\SSK$ in \eqref{SSK}. Following the procedure outlined in Section \eqref{sec:2p1}, we can again extract three integers from this set of advanced equations, which we will denote with a label ``$a$'',
\begin{align}
    E_I^a=0 \then (l_a,g_a,e_a)\,.
\end{align}
As we will discuss later, this duality operation relating $E^r_i$ to $E^a_I$ can also be extended to the case of non-linear equations using the functional antiderivative. 

\paragraph{The number of degrees of freedom} The main result of this work is that the number of degrees of freedom for the set of equations $E_i^r=0$ for the $N_r$ fields $\pr$ is given by the following formula:
\begin{align}
    \text{Open system: } \quad \dof=N_r-\frac12(l_r+g_a+e_a)\,.
\end{align}
A few comments are in order. The first contribution $-l_r/2$ was to be expected, since every constraint appearing in the retarded equation removes the freedom of specifying one initial condition, hence, half of a degree of freedom, for the retarded fields. Conversely, the fact that one should use $g_a$ and $e_a$ instead of $g_r$ and $e_r$ may appear surprising at first. If we used a gauge identity of the retarded equation to propose a gauge transformation $\Delta \phi$, and if we plugged it inside the MSR action, we would realise that it actually does not leave $\SSK$ invariant. The issue is that, upon integration by part, one finds a relation among the \textit{advanced} equations of motion, while the gauge identity was a statement about the \textit{retarded} equations of motion. Instead, the correct retarded gauge transformation follows from an advanced gauge identity. Indeed, imagine one has found the following advanced gauge identity
\begin{align}\label{advgaugeid}
    \G^a=\sum_{I=1}^{N_a} \sum_{m=0}^M\rho^a_{I,m}\left( \frac{d}{dt} \right)^m E_I^a=0\,.
\end{align}
Then it is straightforward to check that the retarded gauge transformation 
\begin{align}\label{gaugetransf}
    \Delta \pr=\sum_{m=0} (-1)^m\left(\frac{d}{dt}\right)^m\left(\rho^a_{Im} \epsilon\right)\,.
\end{align}
leaves the action invariant 
\begin{align}
    \SSK \simeq \int \pa M_{iI} \Delta \pr \simeq \int \Delta \pr E_I^a  \simeq \int \epsilon \, \G^a =0\,.
\end{align}
The slogan to keep in mind is
\begin{align}
    \text{advanced gauge identity}\quad  \leftrightarrow  \quad \text{retarded gauge transformation}\,, \\
    \text{retarded gauge identity}\quad  \leftrightarrow  \quad \text{advanced gauge transformation}\,.
\end{align}
This shows that it is the numbers $g_a$ and $e_a$ that should appear in the correct formula for the number of retarded degrees of freedom. We will verify this in a few example, including the very simple case of gauge-fixed Maxwell's equations around \eqref{eq: EM gauge fixed EoMs}.


\subsection{MSR vs Schwinger-Keldysh formalism}\label{sec2p3}

Let us briefly comment on the relation between the auxiliary MSR functional used above and the
Schwinger--Keldysh path integral. The Martin--Siggia--Rose construction gives a path-integral
representation of classical stochastic dynamics. Given a set of deterministic equations
\begin{equation}
    E_i^r[\phi^r] = 0 \, ,
\end{equation}
one may introduce auxiliary fields \(\phi_i^a\) and write
\begin{equation}
    S_{\rm MSR}[\phi^r,\phi^a]
    :=
    \int dt\, \phi_i^a E_i^r[\phi^r] \, .
\end{equation}
This is indeed the key idea we borrowed for our construction. The integral over \(\phi_i^a\) imposes the equations of motion \(E_i^r=0\). In the presence of
stochastic sources,
\begin{equation}
    E_i^r[\phi^r] = \xi_i \, ,
\end{equation}
with probability functional \(P[\xi]\), averaging over the noise gives
\begin{equation}
    e^{iS_{\rm MSR}}
    =
    \exp\left(i\int dt\,\phi_i^a E_i^r[\phi^r]\right)
    \left\langle
    \exp\left(-i\int dt\,\phi_i^a \xi_i\right)
    \right\rangle_{\xi} .
\end{equation}
For Gaussian noise this produces the familiar quadratic term in the advanced fields,
\[
    iS_{\rm noise}
    =
    -\frac{1}{2}
    \int dt\,dt'\,
    \phi_i^a(t)N_{ij}(t,t')\phi_j^a(t') \, ,
\]
while higher noise cumulants generate higher powers of \(\phi^a\). Thus the MSR functional, which can be written in a path integral with appropriate sources to compute correlators, is
the natural path-integral formulation of classical deterministic or stochastic dynamics.

By contrast, the Schwinger--Keldysh path integral starts from quantum real-time evolution of a
density matrix. For a \textit{closed} system it has the schematic form
\begin{equation}
    Z[J_+,J_-]
    =
    \int {\cal D}\phi_+\,{\cal D}\phi_-\,
    \exp\left\{
        iS[\phi_+] - iS[\phi_-]
        + i\int dt\,\left(J_+\phi_+ - J_-\phi_-\right)
    \right\} .
\end{equation}
The two fields \(\phi_+\) and \(\phi_-\) are the histories on the forward and backward branches of
the closed time contour. It is often useful to pass to the Keldysh basis
\begin{equation}
    \phi^r := \frac{1}{2}(\phi_+ + \phi_-) \, ,
    \qquad
    \phi^a := \phi_+ - \phi_- \, .
\end{equation}
In this basis \(\phi^r\) is the average, or classical, field, while \(\phi^a\) measures the separation
between the two histories.

The connection with MSR appears in a semiclassical limit. Expanding the unitary part of the
Schwinger--Keldysh action at small \(\phi^a\), one finds
\begin{equation}
    S[\phi_+] - S[\phi_-]
    =
    \int dt\, \phi_I^a
    \frac{\delta S[\phi^r]}{\delta \phi_I^r}
    + O\!\left((\phi^a)^3\right) .
\end{equation}
Thus, at leading order in the difference field, the Schwinger--Keldysh path integral contains an
advanced field multiplying the classical equations of motion. For a closed system this gives the
deterministic, noiseless MSR functional.

For an \textit{open} quantum system, one first integrates out environmental degrees of freedom. This produces a Feynman--Vernon influence functional, and the effective Schwinger--Keldysh action
takes the schematic form
\begin{equation}
    S_{\rm eff}[\phi^r,\phi^a]
    =
    \int dt\, \phi_i^a E_i^r[\phi^r]
    + \frac{i}{2}
    \int dt\,dt'\,
    \phi_i^a(t)N_{ij}(t,t')\phi_j^a(t')
    + O\!\left((\phi^a)^3\right) .
\end{equation}
When the dynamics admits a semiclassical regime in which the expansion in \(\phi^a\) is
dominated by the terms linear and quadratic in \(\phi^a\), the Schwinger--Keldysh functional is
equivalent to an MSR functional for a classical Langevin equation. 

The results of this paper should therefore be understood in this classical or semiclassical sense.
Our counting algorithm is formulated directly in terms of the retarded equations \(E_i^r=0\) and
their dual advanced equations obtained from the MSR functional. Strictly speaking, this applies
to classical deterministic or stochastic dynamics. However, whenever an open Schwinger--Keldysh
theory admits a semiclassical limit of the form above, the same retarded/advanced structure is
present. In that regime the advanced field of the Schwinger-Keldysh formalism becomes the MSR
response field, and the counting of retarded degrees of freedom derived here applies to the
corresponding semiclassical equations.

It is important, however, not to identify the two frameworks too strictly. The class of classical
stochastic systems to which the MSR construction applies is, in general, larger than the class of
systems that can be obtained as the semiclassical limit of an open Schwinger--Keldysh path
integral. In the MSR construction one may start from an arbitrary number \(N_a\) of equations
for an arbitrary number \(N_r\) of fields. By contrast, in the Schwinger--Keldysh path integral
the retarded and advanced fields originate, at least before any further manipulation, from the
doubling of the same set of quantum operators. One might therefore expect the semiclassical
Schwinger--Keldysh description to lead naturally to equal numbers of retarded and advanced
variables. The situation is less clear-cut in the presence of constraints and gauge redundancies.
One may choose to fix some retarded and/or advanced gauge freedom, or to integrate out
constrained variables, and these operations need not be implemented in an identical way in the
two sectors. This can lead to effective descriptions in which the numbers of retarded and advanced
variables are no longer manifestly equal, even if the starting point was a standard doubled
Schwinger--Keldysh theory. Fortunately, the validity of our counting does not depend on resolving
this question. Our results apply directly to the broad class of classical equations of motion,
possibly supplemented by stochastic sources, with arbitrary \(N_r\) and \(N_a\). This class may be
larger than the one obtained from semiclassical limits of open quantum systems, but whenever such
a semiclassical Schwinger--Keldysh limit exists, it falls within the same retarded/advanced
structure studied here. A more systematic understanding of the precise overlap between these two
classes would be interesting, but is not needed for the present analysis.


\section{The role of advanced gauge symmetry}\label{sec:3}

In this subsection, we discuss some of the implications of an advanced gauge symmetry, which emerges any time $g_r \neq 0$ (recall that $e_r \geq g_r$, so also $e_r > 0$). We consider exclusively \textit{classical} dynamics, leaving aside any feature of the quantum theory and/or of the full path integral. For recent work in that direction see e.g.~\cite{Kaplanek:2025moq,Kaplanek:2026kpp}. We will see that advanced gauge symmetry plays at least two important roles. First, it gives necessary conditions on how to upgrade a deterministic set of equations to a stochastic set of equations, as encountered in open systems and the Langevin equation. These constraints are simply a generalisation of the noise constraints found in \cite{Salcedo:2024nex}. Second, in some cases, fixing the advanced gauge symmetry gives a procedure to reduce a given set of equations to a ``\textit{minimal form}'', i.e.~a minimal subset of non-redundant, independent equations.  


\subsection{Noise constraints}

Given a set of \textit{deterministic} linear equations of motion $E_i=0$, we want to know what external sources and stochastic noises we can include while respecting the consistency of the theory. We write
\begin{align}
    E_i=J_i\,, \text{ for } i=1,2,\dots, N_a\,,
\end{align}
where $J_i$ represents both the stochastic sources $\xi_i$, which we define to have vanishing expectation value, $\ex{\xi_i}=0$, and external, deterministic sources $j_i\coloneqq \ex{J_i}$. Now assume that a gauge identity $\G$ existed for the deterministic equations of motion $E_i=0$. These can be written in the general form \eqref{gaugeid}, which we report here for convenience
\begin{align}
\G=\sum_{i=1}^{N_a} \sum_{m=0}^M\rho_{i,m}\left( \frac{d}{dt} \right)^m E_i=0\,.    
\end{align}
While this relation is valid off-shell, namely for all possible values of the dynamic and fields, we can choose to go on shell and substitute the solution of the equations of motion, namely $E_i=J_i$. From this, we conclude that on the solutions of the equations of motion for the dynamical fields we must have: 
 \begin{align}\label{noisecontraint}
\sum_{i=1}^{N_a} \sum_{m=0}^M\rho_{i,m}\left( \frac{d}{dt} \right)^m J_i=0\,.    
\end{align}
Simply by taking the expectation value of this expression, we discover that the external deterministic sources $j_i$ as well as the stochastic sources $\xi_i$ have to satisfy this condition separately. These restrictions were dubbed \textit{noise constraints} in \cite{Salcedo:2024nex}. This in turn implies the existence of advanced gauge transformations, as we discussed previously.

It is useful to see this phenomenon in the toy model discussed in the introduction. In the MSR action, any term quadratic in the advanced fields must be invariant under
\eqref{eq:toy-advanced-gauge-transformation}. For a general quadratic noise term
\begin{align}
    S_{\rm noise}
    =
    \frac{i}{2}
    \int dt\,
    \psi_i N_{ij}\psi_j\,,
\end{align}
this requires
\begin{align}
    R^\dagger_i N_{ij}=0\,,
    \qquad
    R_i\epsilon
    =
    \begin{pmatrix}
        \dot\epsilon \\
        -\epsilon \\
        \ddot\epsilon+\epsilon
    \end{pmatrix}\,,
    \label{eq:toy-noise-transversality}
\end{align}
as an operator equation. In particular, if $N_{ij}$ is taken to be a constant algebraic matrix, then
\begin{align}
    -N_{1j}\partial_t
    -
    N_{2j}
    +
    N_{3j}(\partial_t^2+1)
    =
    0
\end{align}
for each $j$, and hence all components of $N_{ij}$ must vanish. Non-trivial local noise terms, if present, have to be built from gauge-invariant combinations of the advanced fields, for example
\begin{align}
    \Psi_1
    &\coloneqq
    \psi_1+\dot\psi_2\,,
    &
    \Psi_2
    &\coloneqq
    \psi_3+\ddot\psi_2+\psi_2\,,
\end{align}
which are invariant under \eqref{eq:toy-advanced-gauge-transformation}.

In summary, we have shown that a given set of deterministic equations of motion $E_i=0$ can be upgraded to a set of stochastic differential equations $E_i=J_i$ if and only if the noise terms $J_i$ satisfy the noise constraint in \eqref{noisecontraint}, where the coefficients $\rho_{Im}$ are determined by the gauge identities of the original set of equations.


\subsection{Fixing the advanced gauge}

As we have seen, advanced gauge transformations are related to gauge identities among the retarded equations of motion. Each gauge identity tells us that, fully off-shell, one of the equations of motion can be written in terms of linear combinations of the others and their time derivatives. This suggests the possibility of reducing the original set of retarded equations to a smaller subset while retaining exactly the same space of solutions. As we now explain, this is indeed possible in some specific cases.


\paragraph{A general property of gauge identities} It turns out that, for every gauge identity $\G$, there is always at least one non-vanishing coefficient $\rho_{I0}$ with $m=0$ for some $I$:
\begin{align}\label{m=0}
    \exists \,, I : \rho_{I,m=0}\neq 0\,.
\end{align}
This can be seen in two ways. If for a given $\G$ one has  $\rho_{I,m=0}=0$ for every $I$, then one can always find other coefficients $\tilde \rho_{I,m}$ such that
\begin{align}
    \tilde \G \coloneqq \sum_{I=1}^{N_r} \sum_{m=0}^{M-1}\tilde \rho_{I,m}\left( \frac{d}{dt} \right)^m E_I\,, \quad \text{ and } \frac{d}{dt} \tilde \G = \G\,.
\end{align}
Then it must be true that off-shell $\dot{\tilde \G}=0$ and hence $\tilde \G = $ const. But since $\tilde \G$ is linear in the fields and this relation must be valid for all possible field configurations, the constant must be zero and we must have  $\tilde \G=0$. If now $\tilde \rho_{I,m=0}\neq 0$ for some $I$ we have proven the claim. Otherwise, if $\tilde \rho_{I,m=0}=0$ for all $I$, we iterate this procedure. Since at least some of the original $\rho_{I,m}$ must be non-vanishing, this procedure must terminate to give a gauge identity with coefficients $\bar\rho_{I,m}$ for which $\bar\rho_{I,m=0}\neq 0$ for at least some value of $I$. Then we should replace the original gauge identity $\G$ with this gauge identity $\bar \G$ for which \eqref{m=0} is true. 

An alternative argument focuses on the gauge transformation associated to the gauge identity $\G$. If the condition \eqref{m=0} is violated, then the associated gauge transformation involves the gauge parameter $\epsilon$ with at least one time derivative. In this case we find another gauge transformation simply by redefining $\tilde \epsilon = \dot \epsilon$. By iterating this procedure, we eventually find a gauge transformation for which \eqref{m=0}. Henceforth, we will assume that \eqref{m=0} is true. 


\paragraph{A reduced set of equations} In the very specific case in which
\[  \rho_{\bar I,m=0}\neq 0 \quad \text{and} \quad \rho_{\bar I,m\neq 0}=0\,, \]
it is possible to reduce the retarded equations of motion to a smaller subset. Indeed, we can re-write the corresponding gauge identity as the off-shell relation
\begin{align}
 \rho_{\bar I ,0}E_{\bar I}=-\sum_{I \neq \bar I}\sum_m \rho_{I,m} \left( \frac{d}{dt} \right)^m E_I\,.
\end{align}
This tells us that the $\bar I$ equation can be written as a linear combination of the other equations and their time derivatives, and therefore can be dropped from the system of differential equations. It is important to notice that this was only possible because the $E_{\bar I}$ equation appeared in the gauge identity without time derivatives. When this is not the case, one cannot drop the equation without losing constraints on the initial conditions. This is, in fact, the same phenomenon that takes place in ordinary gauge theories. There, a generic gauge fixing at the level of the action leads to equations of motion that miss the constrained equation, as for example in the temporal gauge of electromagnetism. In that situation, the correct procedure is to first derive the equations of motion and then gauge fix or add Lagrange multipliers that implement the constraints. We do not attempt a similar construction here. 

The possibility above is exemplified by our toy model. 


\section{Examples}\label{sec:ex}

In this section, we collect a number of examples of our procedure, organised in somewhat increasing amount of complexity. We start with simple two-field models, then discuss electromagnetism and finally gravity. 


\subsection{Two fields, two equations} 

Let us start with a simple toy-model, taken from  \cite{Salcedo:2024nex}, of two pairs of retarded and advanced fields. The action is of the form 
\begin{align}
\SSK = \int \frac{dtd^3k}{(2\pi)^3} \phi^iM_{iJ}\phi^J_r    \,,
\end{align}
where $i, J = \{1,2\}$, the matrix $M_{iJ}$ is
\begin{equation}
    M = \begin{pmatrix}
-\partial_t^2 - k^2 & k^2+ \lambda\partial_t^2  \\
\partial_t^2 + k^2 \lambda & -\partial_t^2 - k^2
\end{pmatrix}\,, \label{paper example matrix M}
\end{equation}
and $\lambda\neq 1$ is some coupling constant. Note that for $\lambda \neq 1$ the matrix $M$ is not Hermitian, and hence the theory is dissipative.
The retarded equations of motions for $\phi^J_r$ are given by
$$E^{(0)}_i \coloneqq \frac{\delta \SSK}{\delta \phi_a^i}=  M_{iJ}\phi^J_r =0\,,$$
and can be expressed in the form:
\begin{equation}
   E^{(0)}_i \coloneqq W^{(0)}_{i J} \ddot \phi^J_r + K^{(0)}_i= \begin{pmatrix}
-1 &\lambda \\
1 & -1
\end{pmatrix} \begin{pmatrix}
    \ddot \phi^1_r\\\ddot \phi^2_r
\end{pmatrix} + 
\begin{pmatrix}
    -k^2 \phi^1_r + k^2 \phi^2_r\\  \lambda k^2 \phi^1_r - k^2 \phi^2_r
\end{pmatrix}\,.
\end{equation}

The matrix $W^{(0)}_{i J}$ has no null-vectors for $\lambda\neq 1$. Consequently, there are no constraints and no gauge identities, and the number of degrees of freedom is 2.\\

\subsubsection*{Case $\lambda=1$}
As noticed before, the case $\lambda=1$ corresponds to Hermitian case, therefore the retarded and advanced sector of the analysis produce the same results.
Setting $\lambda=1$ the matrix $W$ reads
\begin{equation}
    W^{(0)}_{i J}= \begin{pmatrix}
-1 &1 \\
1 & -1
\end{pmatrix}\,, 
\end{equation}
and has the null-vector
\begin{equation}
    w=(1,1)\,,
\end{equation}
which gives the constraint
\begin{equation}
    \C\coloneqq -k^2 \phi^1_r + k^2 \phi^2_r+ k^2 \phi^1_r - k^2 \phi^2_r = 0 \,.
\end{equation}
Since $\C$ vanishes identically, this is actually a gauge identity, and the algorithm terminates.
Let us reformulate it in the form 
\begin{equation}
    \G\coloneqq w\cdot E^{(0)}=E^{(0)}_1+E^{(0)}_2 \equiv0\,,
\end{equation}
from which we can easily extract the gauge transformation:
\begin{equation}
    \begin{cases}
        \delta \phi_r^1=\delta \phi_a^1 = \epsilon(t)\,,\\
        \delta \phi_r^2=\delta \phi_a^2 = \epsilon(t)\,,
    \end{cases}
\end{equation}
where there is only one parameter $\epsilon$ and no dependence on its time derivative, therefore $e_r=e_a=1$.
The number of degrees of degrees of freedom is then
\begin{equation}
    \#d.o.f.=N_r-\frac{1}{2}(l+g+e)=2-\frac{1}{2}(0+1+1)=1\,.
\end{equation}

This simple example shows how sensitive the number of degrees of freedom is to parameter tuning in the equations. 
In this case, a single parameter $\lambda$ was sufficient to transition from two to one degree of freedom, as summarised in the following table.
\begin{table}[h]
\begin{tabular}{|l|l|l|l|l|l|}
\hline
                & Number of fields & Constraints & Gauge Id. & Gauge  Transf. & Number of d.o.f     \\ \hline
$\lambda\neq 1$ & $N=2$             & $l=0$        & $g=0$             & $e=0$              & $2 $ \\ \hline
$\lambda = 1$   & $N=2$             & $l=0$        & $g=1$             & $e=1$              & $ 2- \frac{1}{2}(1+1)=1$ \\ \hline
\end{tabular}
\end{table}


\subsection{A square system with inequivalent retarded and advanced analyses}

It is tempting to expect that, when the number of equations is equal to the number of fields, the retarded and advanced analyses should give identical answers. This is not the case. A simple counterexample is provided by two fields \(\phi\) and \(\sigma\), and two equations,
\begin{align}
E^r_1 &:= (\partial_t+\Omega)(\phi+\sigma)=0\,,
\\
E^r_2 &:= \partial_t(\partial_t+\Omega)(\phi+\sigma)=0\,.
\end{align}
Thus \(N_a=N_r=2\). We can already see by eye that both fields enter the equations, but only through the gauge-invariant combination
\begin{equation}
z:=\phi+\sigma\,.
\end{equation}
The orthogonal combination $w:=\phi-\sigma$ is absent from the equations and is therefore pure gauge. The algorithm will confirm this expectation.

First we note that the first equation is already a constraint. Taking its time derivative we find the gauge identity
\begin{equation}
\partial_t E^r_1-E^r_2\equiv0\,.
\end{equation}
The gauge identity involves one derivative acting on the first equation. The retarded analysis stops here giving
\begin{align}
    l_r&=1\,, & g_r&=1\,, & e_r&=2\,.
\end{align}
The associated auxiliary action is
\begin{equation}
\SSK
=
\int dt\,\left[
a_1 E^r_1+a_2 E^r_2
\right] .
\end{equation}
Using \(E^r_2=\partial_t E^r_1\), this can be written, up to boundary terms, as
\begin{equation}
\SSK
=
\int dt\, (a_1-\dot a_2)(\partial_t+\Omega)(\phi+\sigma)
=
-\int dt\, (\phi+\sigma)(-\partial_t+\Omega)(a_1-\dot a_2)\,.
\end{equation}
The advanced equations are therefore
\begin{align}
E^a_\phi &:=
-(-\partial_t+\Omega)(a_1-\dot a_2)=0\,,
\\
E^a_\sigma &:=
-(-\partial_t+\Omega)(a_1-\dot a_2)=0\,.
\end{align}
They obey the algebraic gauge identity
\begin{equation}
E^a_\phi-E^a_\sigma\equiv0\,.
\end{equation}
Thus the retarded and advanced analyses give different results. For the advanced analysis we find
\begin{align}
    l_a&=1\,, & g_a&=1 \,, & e_a&=1\,.
\end{align}
As expected, the advanced identity generates the retarded gauge transformation
\begin{equation}
\delta \phi=\epsilon\,,
\qquad
\delta \sigma=-\epsilon\,,
\end{equation}
under which
\begin{equation}
\delta z=0\,,
\qquad
\delta w=2\epsilon\,.
\end{equation}
Therefore \(w\) is gauge, while \(z\) is physical. The counting is now immediate:
\begin{align}
    \dof=N_r-\frac{1}{2}(l_r+g_a+e_a)=2- \frac{1}{2}(1+1+1)=\frac{1}{2}\,.
\end{align}
This is indeed confirmed by using the retarded identity, where the only independent equation is
\begin{equation}
(\partial_t+\Omega)z=0\,.
\end{equation}
Its solution is $z_0 e^{-\Omega t}$, and so the physical solution space contains one arbitrary constant. 

This example illustrates why the formula for the number of degrees of freedom must involve the gauge identities of the advanced equations. Moreover, even though \(N_a=N_r\), the retarded and advanced identities have different derivative structure. The advanced identity is the one that detects the retarded gauge redundancy \(\phi\to\phi+\epsilon\), \(\sigma\to\sigma-\epsilon\), and hence correctly removes the unphysical combination \(w=\phi-\sigma\). If one assumed that the retarded and advanced analyses were automatically equivalent whenever \(N_a=N_r\), one would miss precisely this distinction.


\subsection{Electromagnetism}

Electromagnetism is a particularly interesting example in this context. On the one hand, being a gauge theory, it can be used as a well-known model to illustrate how the algorithm works regardless of the choice of gauge fixing. On the other hand, it is itself a theory of interest in the study of open systems, both for its own applications and as a simpler model toward gravity. Systematically computing the number of degrees of freedom and studying the constraint and gauge structure of both the retarded and advanced sectors is complementary to the analysis presented in \cite{Salcedo:2024nex}.

\paragraph{Unitary case, no gauge fixing, no sources.} Therefore, before generalising it to include dissipation, let us begin with the standard, unitary action of Maxwell theory:
\begin{equation}
    S_{EM}=-\frac{1}{4}\int d^4x F_{\mu\nu}F^{\mu\nu}\,.
    \label{eq: EM unitary action}
\end{equation}
Recall that, the theory being unitary, there is no need so far to differentiate between the retarded and the advanced sector of the analysis. We first perform the algorithm starting with the standard action \eqref{eq: EM unitary action} to confirm the expected results, i.e. two propagating degrees of freedom.
The equations of motion are
\begin{equation}
    E^{(0)}_\nu= \partial^\mu F_{\mu\nu}\coloneqq W^{(0)}_{\mu\nu}\ddot A^\nu + K^{(0)}_\mu = \begin{pmatrix}
0 & 0 \\
 0  & -\delta_{ij}
\end{pmatrix} \begin{pmatrix}
    \ddot A^0\\\ddot A^j
\end{pmatrix} + 
\begin{pmatrix}
   +k^2 A^0 -ik_j\dot A^j  \\
    -k^2 A_i -ik_i \dot A^0 + k_i k_jA^j
\end{pmatrix}\,.
\end{equation}
There is clearly one constraint 
\begin{equation}
    \C^{(0)}\coloneqq  +k^2 A^0 -ik_j\dot A^j =0\,,
\end{equation}
given by the null vector $w^{(0)}=(1,\vec 0)$. Adding its time derivative to the equations of motion will give
\begin{equation}
    E^{(1)}_q\coloneqq \begin{pmatrix}
0 & 0 \\
 0  & -\delta_{ij}\\
 0&-ik_j
\end{pmatrix} \begin{pmatrix}
    \ddot A^0\\\ddot A^j
\end{pmatrix} + 
\begin{pmatrix}
   k^2 A^0 -ik_j\dot A^j  \\
    -k^2 A_i -ik_i \dot A^0 + k_i k_jA^j\\
    k^2 \dot A^0
\end{pmatrix}\,.
\end{equation}
The rectangular matrix $W^{(1)}_{qj}$, where $q=0,...,4$ has the null vector
\begin{equation}
    w^{(1)}=(0, -i\ k^i, 1)\,,
\end{equation}
that contracted with $K_q^{(1)}$ gives 
\begin{equation}
    w^{(1)}\cdot K^{(1)}= i\ k^2 k^iA_i - k^2 \dot A^0 -i k^2 k_j A^j +k^2 \dot A^0 = 0\,.
\end{equation}
Since this equation vanishes identically, we conclude that we have found a gauge identity and the algorithm stops. Let us reformulate it in the form
\begin{equation}
        \G\coloneqq w^{(1)}\cdot K^{(1)}=-i\ k^iE^{(0)}_i+\dot E^{(0)}_0=0\,.
\end{equation}
From here the standard gauge transformation follows, $A^\mu\to A^\mu+ \partial^\mu \epsilon $ and we see that $e=2$ since the transformation depends also on the first time derivative of $\epsilon$. As we expected we found a gauge symmetry and, having found $l=1$, $g=1$ and $e=2$, the number of degrees of freedom are
\begin{equation}
    \# \text{d.o.f.}=4-\frac{1}{2}(1+1+2)=2\,,
\end{equation}
as expected.


\paragraph{Coulomb gauge fixing in the action}
Having established the proper functioning of the procedure, we now investigate the effect of gauge fixing, expecting, as required, the same number of physical degrees of freedom. Here we have two options: the first consists in performing the gauge fixing at the action level and deriving the equations of motion from the resulting gauge-fixed action. The second option consists in gauge fixing the equations of motion after having obtained them from a non-gauge-fixed action. We analyse these two cases separately to show the need for a generalisation of the algorithm in order to obtain a consistent result. We will work using the Coulomb gauge, i.e.
\begin{equation}
    \partial_i A^i=0\,.
\end{equation}
Recall that we can always choose the reference frame such that the propagation direction is proportional to $k_3 $ , i.e.
\begin{equation}
    k^i=(0,0,k)\,.
\end{equation}
With this choice Coulomb gauge becomes 
\begin{equation}
    k_iA^i=0 \implies k_3A^3=0 \implies A^3=0\,.
\end{equation}
Then $N = \text{number of fields} = 3$, and the action has the simplified form
\begin{equation}
    S_{\text{gauge fixed}}=\int \frac{dt d^3k}{(2\pi)^3} \sum_{i=1,2} \frac{1}{2}(\dot A^{i})^2-\frac{1}{2}k^2 (A^i)^2 - \frac{1}{2} k^2 (A^0)^2\,,
\end{equation}
from which the equations of motions are
\begin{equation}
    \begin{cases}
        k^2A^0=0\,,\\
        \ddot A^1 +k^2 A^1=0\,,\\
        \ddot A^2 +k^2 A^2=0 \,,      
    \end{cases}
\end{equation}
where it is clear that the only constraints are 
\begin{equation}
    \C^{(0)}\coloneqq k^2A^0=0\,,
\end{equation}
and its time derivative
\begin{equation}
    \C^{(1)}\coloneqq k^2\dot A^0=0\,,
\end{equation}
after which the algorithm ends. This analysis works well: we have found two constraints ($l=2$) and no gauge identity, as we expect after having fixed it. The number of degrees of freedom is
\begin{equation}
    \#d.o.f.=N-\frac{1}{2}(l+g+e)=3-\frac{1}{2}(2)=2\,.
\end{equation}


\paragraph{Coulomb gauge fixing in the equations of motion} In this second case we first derived the equations of motion from the non-gauge-fixed action \eqref{eq: EM unitary action}, which, choosing again $k^i=(0,0,k)$, take the form:
\begin{equation}
    \begin{cases}
        k^2A^0-ik\dot A^3=0\,, \\
        - \ddot A^1 -k^2A^1=0\,, \\
        - \ddot A^2 -k^2A^2=0\,,\\
        - \ddot A^3  -i k \dot A^0 =0\,.\\
    \end{cases}
\end{equation}
Fixing the gauge as before means setting $A^3=0$, and the equations of motion then become
\begin{equation}
    \begin{cases}
        k^2A^0=0\,,\\
        - \ddot A^1 -k^2A^1=0\,,\\
        - \ddot A^2 -k^2A^2=0\,,\\
        -i k \dot A^0 =0\,.\\
    
    \end{cases}
    \label{eq: EM gauge fixed EoMs}
\end{equation}
This is exactly the case where the set of equations, as given, cannot possibly derive from an action, since we have four equations but only three fields, so $N_r=3 \neq N_a=4$. Formally this set of equations does not derive from gauge fixing the action in \eqref{eq: EM unitary action}. \\
In this case there are already two constraints in the 0th step, namely
\begin{equation}
    \begin{cases}
        \C^{(0)}_1\coloneqq k^2A^0=0\,,\\
       \C^{(0)}_2\coloneqq -i k \dot A^0 =0\,.\\
    \end{cases}
\end{equation}
Applying the algorithm in section \eqref{sec:2p1} then gives a gauge identity at the first step corresponding to 
\begin{equation}
    \G=\dot E_1 - i k E_4=0\,.
    \label{eq: EM gauge identity advanced}
\end{equation}
The algorithm then terminates because differentiating the second constraint does not lead to any new left null eigenvectors. Therefore we find
\begin{align}
    l_r & =2\,, & g_r&=1\,, & e_r&=2\,.
\end{align}
If we blindly applied the results for unitary theory, we would find two inconsistencies. First, there should be no remaining gauge symmetry since we imposed Coulomb gauge and yet we have found $e_r=2$, which in a unitary theory gives the gauge invariance discussed around \eqref{unitaryaction}. Second, using the \textit{unitary} number of degrees of freedom formula in \eqref{dofunitary} would give the absurd result
\begin{align}
    \dof \overset{??}{=}N_r-\frac12(l_r+g_r+e_r)=3-\frac12(2+1+2)=\frac12\,.
\end{align}
The solution, as anticipated, is that the gauge identity \eqref{eq: EM gauge identity advanced} is associated with a gauge transformation in the advanced sector and therefore should not be counted in the computation of the physical number of degrees of freedom. Note that having a remaining gauge symmetry is perfectly allowed in the advanced sector, as we have performed the gauge fixing on the retarded fields only.\\

To check the proper functioning of the generalised algorithm, we perform the analysis of the advanced sector to confirm the absence of retarded gauges. To do so we construct the MSR action by multiplying each equation in \eqref{eq: EM gauge fixed EoMs} by an advanced field.
\begin{equation}
    \SSK=\int  \frac{dt d^3k}{(2\pi)^3} a^\mu E^{(0)}_{GF\mu}\,.
\end{equation}
After performing integration by parts we obtain the equation of motion for the advanced fields:
\begin{equation}
    \begin{cases}
        k^2 a^0+i\ k\ \dot a^3=0\,,\\
        -\ddot a^1 -k^2 a^1=0\,,\\
        -\ddot a^2 -k^2 a^2=0\,,\\
    \end{cases}
\end{equation}
where it is clear that the only constraint is 
\begin{equation}
    \C^{(0)}\coloneqq k^2a^0 +i\ k\ \dot a^3=0\,,
\end{equation}
after which the algorithm ends, confirming the absence of retarded gauge identities. Therefore we find
\begin{align}
    l_a&=1\,, & g_a&=e_a=0\,.
\end{align}

The physical degrees of freedom are again
\begin{equation}
    \dof=N_r-\frac{1}{2}(l_r+g_a+e_a)=3-\frac{1}{2}(2)=2\,.
\end{equation}
as expected.


\paragraph{Including dissipation: light in a medium}
After studying the unitary Maxwell theory, we can generalise it by including dissipation. In order to do this we rewrite the action in the MSR formalism.
\begin{align}
    \SSK = \int dt \int \frac{d^3k}{(2\pi)^3} \, a^\mu M_{\mu\nu} A^\nu\,.
\end{align}
where $A^\mu$ are the retarded fields,  $a^\mu$  the advanced fields and $\mu$ is the spacetime index. The dissipative action takes the form

\begin{equation}
	M_{\mu\nu} =
	\begin{pmatrix}
		k^2 & - i \partial_0 k_j \\[5pt]
		- i \gamma_2 k_i & 
		 \gamma_2 \partial_0  \delta_{ij} + \gamma_3 (k^2 \delta_{ij} - k_i k_j) - 2i \gamma_4 \epsilon_{ijl} k_l
	\end{pmatrix}\,,
\end{equation} 
from \cite{Salcedo:2024nex}. The coefficients $\gamma_2$,$ \gamma_3$ and $\gamma_4$ can be expanded at different orders of $\omega$:
\begin{equation}\label{LiaM coefficients expansion}
	\gamma_i=\Gamma_i-ig_i\omega\ + \overset{\sim}{g}_i \omega^2\,.
\end{equation}
Since we consider terms up to second order in time derivatives, the expansion of $\gamma_2$ terminates at the second term:
$$\gamma_2=\Gamma_2-ig_2\omega\,.$$
The retarded sector analysis produces one constraint ($l_r=1$) and one gauge identity ($g_r=1$)
\begin{align}
&\C_r^{(0)}\coloneqq k^2 A^0-i k_i\dot A^i =0\,,\\
&\G_r \coloneqq \frac{\Gamma_2}{g_2}\dot \C_r^{(0)}+C_r^{(1)} = \frac{\Gamma_2}{g_2} E^{(0)}_0 - \frac{i k^i}{g_2} E^{(0)}_i+ \dot E^{(0)}_0 \equiv 0\,.
\end{align}
From the gauge identity one can extract the parameter $e_r=2$ since there is a dependency on the first derivative of the original equations of motions.
The advanced analysis also produces one constraint ($l_a=1$) and one gauge identity ($g_a=1$):
\begin{align}
	&\C_a^{(0)}\coloneqq k^2 a^0+i g_2 k_i\dot a^i -i \Gamma_2 k^ia_i =0 ,\\
	&\G_a \coloneqq C_r^{(1)} = i k^i E^{(0)}_i + \dot E^{(0)}_0 \equiv 0\,.
\end{align}
From this advanced gauge identity we deduce, using  \eqref{gaugetransf}, the following gauge transformations
\begin{align}
	\delta A^0 &=-\dot \epsilon\,, & \delta A^i&= i k^i \epsilon\,.
\end{align}
From here we can extract the parameter $e_a =2$. Moreover this gauge transformation corresponds to 
\begin{eqnarray}
	A_\mu \longrightarrow A_\mu + \partial_\mu \epsilon\,,
\end{eqnarray}
meaning that dissipation does not modify the gauge transformations for the retarded fields. Summarizing
\begin{align}
    l_r&=1\,, & g_r&=1\,, & e_r&=2\,, \\
    l_a&=1\,, & g_a&=1\,, & e_a&=2\,.
\end{align}
Therefore, combining the retarded and the advanced analysis we confirm that the number of degrees of freedom is
\begin{eqnarray}
    \#d.o.f.=N_r-\frac{1}{2}(l_r+g_a+e_a)=4-\frac{1}{2}(1+1+2)=2\,.
\end{eqnarray}


\paragraph{Massive case: Proca Theory}
One may also consider another instructive extension of Maxwell theory, namely the inclusion of a mass term. The corresponding Proca action is
\begin{equation}
	S_{\text{Proca}}=\int d^4x(-\frac{1}{4} F_{\mu\nu}F^{\mu\nu}-\frac{1}{2}m^2 A_\mu A^\mu)\,.
	\label{eq: Proca unitary action}
\end{equation}
The presence of the mass term explicitly breaks gauge invariance. Therefore we expect three propagating degrees of freedom and two real constraints. Indeed we find
\begin{eqnarray}
\C ^{(1)}\coloneqq m^2(\dot A^0 + i\ k_j A^j) =0\,.
\end{eqnarray}
Note that this vanishes off-shell for $m=0$, indicating the emergence of a gauge identity and therefore a change in the number of degrees of freedom. A detailed study shows that the same counting applies for the generalisation of Proca theory to the non unitary, dissipative case. This will be discussed elsewhere. 



\subsection{General relativity with a perfect fluid}
Having explored electromagnetism in depth, we now turn to general relativity.
We begin with a simple and representative case, namely a perfect fluid, whose dynamics is characterised by an energy–momentum tensor of the form
\begin{equation}
	T_{\mu\nu} = (\rho + p) u_\mu u_\nu + g_{\mu\nu} p \, .
\end{equation}

We study the dynamics of linear perturbations around a flat Friedmann–Robertson–Walker (FRW) background. 
The perturbed metric is written as
\begin{equation}
	g_{\mu\nu}(t,\textbf{x}) = \bar{g}_{\mu\nu}(t) + h_{\mu\nu}(t,\textbf{x}) \, ,
\end{equation}
where $\bar{g}_{\mu\nu}$ is the background metric,
\begin{equation}
	\bar{g}_{\mu\nu} =
	\begin{pmatrix}
		-1 & 0 \\
		0 & a^2(t)\delta_{ij}
	\end{pmatrix} \, ,
\end{equation}
and $h_{\mu\nu}$ denotes a small perturbation.

The background energy–momentum tensor
$\bar T_{\mu\nu} = (\bar \rho + \bar p)\bar u_\mu \bar u_\nu + \bar g_{\mu\nu}\bar p$
takes the form
\begin{equation}
	\bar{T}_{\mu\nu} =
	\begin{pmatrix}
		\bar \rho & 0 \\
		0 & a^2(t)\bar p \delta_{ij}
	\end{pmatrix} \, ,
\end{equation}
in the rest frame of the fluid, where $\bar u_i = 0$ and $\bar u_0 = -1$.

The Einstein equations for this background reduce to the Friedmann equations,
\begin{equation}
	3 M^2 \left(\frac{\dot a(t)}{a(t)}\right)^2 = \bar \rho(t) \, ,
\end{equation}
where $M^2=\frac{1}{8\pi G}$ and
\begin{equation}
	\frac{\ddot a(t)}{a(t)} =
	- \frac{1}{6M^2}(\bar \rho + 3 \bar p)
	= - \frac{1}{6M^2}(1 + 3 w)\bar \rho \, .
\end{equation}
In the last equality we assumed the equation of state $ p = w  \rho$.
The perturbed energy-momentum tensor takes the form
\begin{eqnarray}
	\begin{aligned}
		&\delta T_{00}=-\bar \rho h_{00}+\delta \rho\, ,\\
		&\delta T_{0i}=\bar p h_{0i}-(\bar \rho +\bar p)\delta u_i=w \bar \rho h_{0i}-(1+w)\bar \rho \delta u_i \, , \\
		&\delta T_{ij}= \bar p h_{ij}+a^2 \delta_{ij}\delta p =  w\bar \rho h_{ij}+a^2 \delta_{ij} w \delta \rho \,.
	\end{aligned}
\end{eqnarray}
It is now convenient to introduce the Scalar-Vector-Tensor decomposition (SVT) and study the dynamics of scalars and tensors separately as they do not mix at linear level.
Adopting the notation of \cite{Weinberg:2008zzc}, the fluctuations of the metric are parametrised by 
\begin{align}
	h_{00}(t,\mathbf{x}) &= -E(t,\mathbf{x}), \\[6pt]
	h_{0i}(t,\mathbf{x}) &= a\left[\partial_i F(t,\mathbf{x}) + G_i(t,\mathbf{x})\right], \\[6pt]
	h_{ij}(t,\mathbf{x}) &= a^2 \Big[
	A(t,\mathbf{x})\delta_{ij}
	+ \partial_i \partial_j B(t,\mathbf{x})
	+ \partial_i C_j(t,\mathbf{x})
	+ \partial_j C_i(t,\mathbf{x})
	+ D_{ij}(t,\mathbf{x})
	\Big]\, ,
\end{align}
where $A,\ B,\ E,\ F$ are scalar modes, $C_i,\ G_i$ are vector modes and $D_{ij}$ are tensor modes satisfying
\begin{eqnarray}
	\partial^iC_i=\partial^iG_i=\partial^iD_{ij}=0, \quad \delta^{ij}D_{ij}=0\, .
\end{eqnarray}
The fluctuations of the energy-momentum tensor are parametrised by
\begin{align}
	\delta T_{00} &= -\bar{\rho}\, h_{00} + \delta\rho, \\[6pt]
	\delta T_{0i} &= \bar{p}\, h_{0i}
	- (\bar{\rho} + \bar{p})\left(\partial_i \delta u + \delta u_i^{V}\right), \\[6pt]
	\delta T_{ij} &= \bar{p}\, h_{ij}
	+ a^2 \left(
	\delta_{ij}\,\delta p
	+ \partial_i \partial_j \pi^{S}
	+ \partial_i \pi_j^{V}
	+ \partial_j \pi_i^{V}
	+ \pi_{ij}^{T}
	\right)\, ,
\end{align}
where $\partial_i \pi_i^{V}$, $ \pi_{ij}^{T}$ and $\delta u_i^V$ satisfy
\begin{eqnarray}
	\partial^i \delta u_i^V=\partial^i \pi_i^{V}= \partial^i\pi_{ij}^{T}=0\, , \quad  \delta^{ij}\pi_{ij}^{T}=0\, .
\end{eqnarray}
For a perfect fluid we set 
\[ \pi^{S}=\pi_j^{V}=\pi_{ij}^{T}=0\, .\]
With this formalism the three sectors are decoupled and each of them is described by the corresponding Einstein equations. We will therefore present their analysis separately.


\paragraph{Scalar sector}
In the scalar sector there are seven fields, but they can be reduced to six by making use of the equation of state $p=w \rho$. Therefore the six retarded fields are 
\[\phi_r=\{A,\, B,\,E,\,F,\, \delta \rho,\, \delta u\}\,.\]
The equations of motion for these modes, corresponding in our language to the retarded equations of motion, are taken from \cite{Weinberg:2008zzc}.
\[
\begin{aligned}
& \tfrac{1}{2} a \dot a \dot E + (2\dot a^{2} + a \ddot a) E
+ \tfrac{1}{2} \nabla^{2} A - \tfrac{1}{2} a^{2} \ddot A
- 3 a \dot a \dot A - \tfrac{1}{2} a \dot a \nabla^{2} B
+ \dot a \nabla^{2} F=-\frac{a^2}{2M^2}\left(1 - w \right)\delta\rho\, ,\\
& - \dot a\, \partial_j E + a\, \partial_j \dot A =\frac{a}{M^2} (\bar\rho + \bar p)\, \partial_j \delta u
\, ,\\
&- \frac{1}{2 a^{2}} \nabla^{2} E
- \frac{3 \dot a}{2 a} \dot E
- \frac{1}{a} \nabla^{2} \dot F
- \frac{\dot a}{a^{2}} \nabla^{2} F
+ \frac{3}{2} \ddot A
+ \frac{3 \dot a}{a} \dot A
- \frac{3 \ddot a}{a} E
+ \tfrac{1}{2} \nabla^{2} \ddot B
+ \frac{\dot a}{a} \nabla^{2} \dot B=-\frac{1}{2M^2} \left(\delta\rho + 3\delta p \right) \, ,\\
&\partial_j \partial_k
\left[ E + A - a^{2} \ddot B - 3 a \dot a \dot B
+ 2 a \dot F + 4 \dot a F
\right]
= 0 \, .
\end{aligned}
\label{eq: Weinberg Eoms GR perfect fluid}
\]
The retarded analysis produces four constraints and no gauge identities, so 
\begin{align}
l_r&=4\,, & g_r&=e_r=0 \,.     
\end{align} 
A comment is in order. The absence of retarded gauge identities, namely $g_r=0$, arises because we are only considering the Einstein equations, but we have not added the equations for the conservation of the energy-momentum tensor, which follow from them. For example, had we written schematically 
\begin{align}
    G_{\mu\nu}-T_{\mu\nu}=0 \quad \text{ and } \quad \nabla^\mu T_{\mu\nu}=0\,,
\end{align}
then we would have had the gauge identity
\begin{align}
    \nabla^\mu (G_{\mu\nu}-T_{\mu\nu})-\nabla^\mu T_{\mu\nu}=0
\end{align}
being valid off-shell. A powerful feature of our algorithm is that it correctly counts the number of degrees of free whether or not we include additional dependent equations, such as $\nabla^\mu T_{\mu\nu}=0$ in this case. 

As we explained, the gauge structure is to be found in the analysis of the advanced equations of motion, obtained by the variation of the MSR action with respect to the retarded fields. We construct the MSR action by multiplying each equation in \eqref{eq: Weinberg Eoms GR perfect fluid} by an advanced field
\begin{equation}
	\SSK=\int  \frac{dt d^3k}{(2\pi)^3} \phi_a^i(t,-\mathbf{k}) E^{(0)}_{i}(t,\mathbf{k})\,.
\end{equation}
Note that, since $g_r=e_r=0$, this functional is not invariant under any gauge transformation of the advanced fields. This would be different had we included $\nabla^\mu T_{\mu\nu}$ in the original system of equations. 

After performing integration by parts and having obtained the equation of motion for the advanced fields, we can perform the advanced analysis. From the advanced analysis we found six constraints and two gauge identities:
\begin{equation}
\G^1\coloneqq \tilde{E}^{(0)}_1(t)- \frac{a(t)}{\dot a(t)}\dot{\tilde{E}}^{(0)}_3(t)- \frac{\tilde{E}^{(0)}_{4}(t)}{2\dot a(t)} - \frac{3}{2}(1+w)\tilde{E}^{(0)}_{5}(t)\bar\rho_b(t) - \frac{a(t)}{2\dot a(t)} \tilde{E}^{(0)}_{6}(t)\, ,
\end{equation}
and
\begin{equation}
	\G^2\coloneqq\tilde{E}^{(0)}_{2}(t)
	- \frac{a(t)}{2} \dot{\tilde{E}}^{(0)}_{4}(t)
	- \frac{\dot a(t)}{2} \tilde{E}^{(0)}_{4}(t)\, .
\end{equation}
As it clear by twiddles on the equations $\tilde E_I$, these gauge identities relate the advanced equations of motion. Indeed, as we remarked earlier, there are no gauge identities relating the retarded equations of motion in this example. 

Advanced gauge identities imply gauge transformations of the retarded fields:
\begin{align}
	A &\;\longrightarrow\; A + \varepsilon_1 , & E &\;\longrightarrow\; E + \frac{\dot{\varepsilon_1}}{H} + \frac{3}{2}(1+w)\,\varepsilon_1 , \\
	F &\;\longrightarrow\; F - \frac{\varepsilon_1}{2\dot a} , &
	\delta \rho &\;\longrightarrow\; \delta \rho - \frac{3}{2}(1+w)\,\bar \rho\, \varepsilon_1 , \\
	\delta u &\;\longrightarrow\; \delta u - \frac{1}{2H}\,\varepsilon_1 \,,
\end{align}
for the first gauge identity, and
\begin{align}
	B &\;\longrightarrow\; B + \varepsilon_2, & F &\;\longrightarrow\; F + \frac{a}{2}\,\dot \varepsilon_2\, ,
\end{align}
for the second one.
Each of them contains a dependence on the first derivative of the parameters $\varepsilon_1$ and $\varepsilon_2$, therefore the total number of gauge parameters is $e_a=4 $. Note that redefining
\[\varepsilon_1=2\frac{\dot a}{a}\epsilon_0\, , \quad \varepsilon_2=-\frac{2}{a^2}\epsilon^S\,,\]
one can recover the same expression for the gauge transformations in \cite{Weinberg:2008zzc}.
Summarising, the outputs of the advanced analysis are
\[l_a=6\, , \quad g_a=2\, , \quad e_a=4\,.\]
Therefore, the total number of scalar degrees of freedom is one as expected
\[\#d.o.f.=N_r-\frac{1}{2}(l_r+g_a+e_a)=6-\frac{1}{2}(4+2+4)=1\, .\]
\paragraph{Tensor sector}
The tensor sector is less interesting than the scalar sector because, for a perfect fluid, the corresponding equations reduce to
\begin{align}
    \ddot D_{ij}+3H \dot D_{ij}- \frac{\nabla^2}{a^2} D_{ij} = 0 \, ,
\label{eq: EoM tensor sector GR perfect fluid}
\end{align}
i.e., free wave equations. Within the SVT decomposition, the tensor perturbation $D_{ij}$ is defined as the traceless and transverse part of the metric perturbation. Consequently, it contains only two independent degrees of freedom, each obeying the wave equation \eqref{eq: EoM tensor sector GR perfect fluid}. As expected there are no constraints, no gauge identities and no gauge symmetries.


\subsection{Effective Field Theory of Inflation}

We now move on to the application of this framework to the study the Effective Field Theory of Inflation (EFToI) and its extensions including dissipative and stochastic effects \cite{LopezNacir:2011kk,Salcedo:2024smn,Salcedo:2026sdn}. In this subsection, as a warm up we start with the universal action for for the EFToI. \\



We will work in unitary gauge, where the inflaton is absorbed into the scalar perturbation of the metric. As a consequence, the inflaton becomes homogeneous and defines a natural spacetime foliation. This foliation is characterised by a future-pointing unit vector $n_\mu$, perpendicular to the constant-field hypersurfaces,
\begin{align}n_\mu \coloneqq -\frac{\partial_\mu \phi}{\sqrt[]{-g^{\mu\nu}\partial_\mu\phi\partial_\nu\phi}}\,.
\end{align}
In unitary gauge this becomes \begin{align} n_\mu \dot = - \frac{\delta^0_\mu}{\sqrt[]{-g^{00}}}\,,\end{align}
where the notation $\dot =$ indicates an equality that holds in unitary gauge.

Before turning to the analysis of the open case, we first review the closed EFToI to confirm the expected number of degrees of freedom and to highlight the symmetries of the system.


\subsubsection{Universal part of the Effective Field Theory of Inflation} 
The universal action for the EFToI in unitary gauge takes the form
\begin{align}
S_{\mathrm{univ}} = \int d^4x \sqrt{-g} \left[ \frac{M_{Pl}^2}{2} R - \Lambda(t) - c(t)\, g^{00} \right]\,.
\end{align}
Introducing the advanced fields $a^{\mu\nu}$, this expression can be rewritten in the Keldysh basis:
\begin{align}
\SSK^{\mathrm{univ}} = \frac{1}{2} \int d^4x \sqrt{-g} \left[ M_{Pl}^2 G_{\mu\nu} a^{\mu\nu} + \Lambda(t) g_{\mu\nu} a^{\mu\nu} + c(t) g^{00} g_{\mu\nu} a^{\mu\nu} - 2 c(t)\, a^{00} \right]\,,
\label{eq: universal action EFToI Keldysh basis}
\end{align}
from which it is easy to extract the retarded equations of motion by varying the action with respect to the advanced fields
\begin{align}
	\frac{\delta \SSK^{\mathrm{univ}} }{\delta a^{\mu\nu}}=E_{\mu\nu}^r\propto M_{Pl}^2 G_{\mu\nu} + \Lambda(t) g_{\mu\nu}+ c(t) g^{00} g_{\mu\nu} - 2 c(t)\delta^0_\mu \delta^0_\nu\,, 
\end{align}
The corresponding energy-momentum tensor is
\begin{equation}
    T_{\mu\nu}=-\Lambda(t) g_{\mu\nu}- c(t) g^{00} g_{\mu\nu} + 2 c(t)\delta^0_\mu \delta^0_\nu\,,
\end{equation}
and its background value $\bar{T}_{\mu\nu}$ takes the form
\begin{equation}
	\bar{T}_{\mu\nu} =
	\begin{pmatrix}
		\Lambda+c & 0 \\
		0 & a^2(c-\Lambda) \delta_{ij}
	\end{pmatrix} \, ,
\end{equation}
and gives the two background equations
\begin{equation}
	\begin{cases}
	\left(\frac{\dot a}{a}\right)^2=\frac{1}{3M_{Pl}^2}(\Lambda+c)\,,\\
	\frac{\ddot a}{a}=\frac{1}{3 M_{Pl}^2}(\Lambda-2c)\,.
	\end{cases}
\end{equation}

The perturbed energy-momentum tensor takes the form
\begin{equation}
	\begin{cases}
		\delta T_{00}=-\Lambda h_{00}\,,\\
		\delta T_{0i}=(c-\Lambda) h_{0i}\,,\\
		\delta T_{ij}=(c-\Lambda) h_{ij}+c\  h_{00} a^2\delta_{ij}\,,\\
	\end{cases}	
\end{equation}
As in the previous example we introduce the Scalar-Vector-Tensor decomposition (SVT) to parametrise the fluctuations of the metric, and again we focus on the scalar sector, namely on the fields $A,B,E,F$. The expression for $\delta G_{\mu\nu}$, taken from \cite{Flauger:2009uta}, are given by
 \begin{align}
	\delta G_{00} &=
	-\frac{\nabla^2}{a^2} A
	+ 3\frac{\dot a}{a}\,\dot A
	+ \frac{\dot a}{a}\nabla^2 \dot B
	- \frac{2\dot a}{a^2}\nabla^2 F \,, \label{delta G_00}
	\\[6pt]
	\delta G_{0i} &=
	\partial_i\!\left(
	\frac{\dot a}{a}E - \dot A
	- \left(\frac{\dot a^2}{a^2} + 2\frac{\ddot a}{a}\right)aF
	\right)+...\,, \label{delta G_0i}
	\\[6pt]
	\delta G_{ij} &=
	\delta_{ij}\bigg(
	\dot a^2\,E + 2a\ddot a\, E + \frac{1}{2}\nabla^2 E + a\dot a\dot  E -(\dot a^2 + 2a\ddot a)A \nonumber
	\\
	&\qquad+ \frac{1}{2}\nabla ^2 A
	- a^2\ddot A
	- 3a\dot a \dot A - \frac{1}{2}a^2 \nabla^2\ddot B
	- \frac{3}{2}a\dot a \nabla^2 \dot B
	+2\dot a \nabla^2 F
	+ a \nabla^2 \dot F \bigg) \nonumber
	\\
	&\quad
	+ \partial_i\partial_j\!\left(
	-\frac{1}{2}E - \frac{1}{2}A
	- (\dot a^2 + 2a\ddot a)B
	+ \frac{1}{2}a^2\ddot B
	+ \frac{3}{2}a\dot a \dot B
	- 2\dot a F - a\dot F
	\right) + ... \,,
	\label{delta G_ij}
\end{align}
where the dots denote contributions from vector and tensor modes.
Note that \eqref{delta G_ij} is split into two independent parts ($\sim \delta_{ij}$ and $\sim \partial_i\partial_j$), which give rise to two separate equations.
The (retarded) equations of motion for the scalar sector then take the form
\begin{equation}
		\begin{aligned}
		\E_1\coloneqq&\frac{k^2}{a^2} A + 3 \frac{\dot a}{a} \dot A - \frac{\dot a}{a} k^2 \dot B + 2 \frac{\dot a}{a^2} k^2 F
		- \frac{\Lambda}{M_{\mathrm{Pl}}^2} E=0\,,
		\\[6pt]
		\E_2\coloneqq&\frac{\dot a}{a} E - \dot A - \left[\left(\frac{\dot a}{a}\right)^2 + 2 \frac{\ddot a}{a}\right] a F
		- \frac{c - \Lambda}{M_{\mathrm{Pl}}^2} a F=0\,,
		\\[6pt]
		\E_3\coloneqq&\dot a^2 E + 2 a \ddot a\, E - \frac{k^2}{2} E + a \dot a\, \dot E
		- (\dot a^2 + 2 a \ddot a) A - \frac{k^2}{2} A - a^2 \ddot A - 3 a \dot a\, \dot A
		\\
		&\quad\quad\quad + \frac{a^2 k^2}{2} \ddot B + \frac{3}{2} a \dot a\, k^2 \dot B
		- 2 \dot a\, k^2 F -a k^2 \dot F
		 +\frac{a^2}{M_{\mathrm{Pl}}^2} \left[(\Lambda - c) A + c E \right]=0
		\,,\\[6pt]
		\E_4\coloneqq&-\frac{1}{2} E - \frac{1}{2} A - (\dot a^2 + 2 a \ddot a) B
		+ \frac{a^2}{2} \ddot B + \frac{3}{2} a \dot a\, \dot B
		- 2 \dot a\, F - a \dot F
		- \frac{c - \Lambda}{M_{\mathrm{Pl}}^2} a^2 B=0 \,.
	\end{aligned}
	\label{eq: retarded eoms EFToI closed}
\end{equation}
The retarded analysis produces three constraints and one gauge identity, so 
\begin{align} 
l_r&=3\,, & g_r&=1\,, & e_r&=2\,.
\end{align}
We can now perform the advanced analysis by varying the action \eqref{eq: universal action EFToI Keldysh basis} with respect to the advanced fields in order to obtain the advanced equations of motion $\tilde{E}_{\mu\nu}^a$. The advanced analysis also produces three constraints, which we omit to write explicitly, and one gauge identity: 
\begin{equation}
	\G^1\coloneqq \frac{\dot a}{a}\tilde{\E}_4-\frac{2}{a}\tilde{\E}_2+\dot{\tilde{\E}}_4\equiv 0\, ,
\end{equation}
with the corresponding gauge transformation:
\begin{align}
	B &\;\longrightarrow\; B - \frac{2}{a} \varepsilon, & F &\;\longrightarrow\; F +\frac{\dot a}{a}\varepsilon -\dot \varepsilon\, .
    \label{gauge transformation EFToI}
\end{align}
This results in the advanced counting
\begin{align}
    l_a&=3 \,, & g_a&=1 \,, & e_a&=2 \,.
\end{align}
Note that this is different from the counting we found in the case of a single perfect fluid. However, the total number of scalar degrees of freedom is again one, as expected, because the smaller number of gauge identities and gauge transformation in the EFT of Inflation, as compared to the perfect fluid, precisely compensates for the smaller number of fields:
\begin{align}
\#d.o.f.=N_r-\frac{1}{2}(l_r+g_a+e_a)=4-\frac{1}{2}(3+1+2)=1\, .
\end{align}
This is of course the expected result and provides a non-trivial check of the validity of our algorithm. Next, we move to dissipative modifications. 


\subsection{Effective Field Theory of Dark Energy}

We now turn to extensions of the universal part. In \cite{Salcedo:2025ezu} a theory of dissipative gravity in the presence of an unknown medium was proposed. It was soon realised that this construction missed some important constraints. In \cite{Christodoulidis:2025vxz}, Christodoulidis and Gong showed that choosing generic operators in the Schwinger-Keldish (SK) action leads to scalar equations of motions that admit only trivial solutions where all scalars vanish. Then they showed that some specific tuning of operators can be chosen to find consistent scalar equations with non-trivial solutions. Their modified Einstein equations read
\begin{align} \label{Perseas}
G_{\mu\nu}
+
\left[
 \Lambda
-\frac{1}{2}  c \,\delta g^{00}
-\frac{1}{4}M^2\left(\delta g^{00}\right)^2
\right] g_{\mu\nu}+&
\left(  c +M^2 \delta g^{00}\right)\delta^0_\mu \delta^0_\nu\\ \nonumber
&\qquad +
\Gamma\left(K_{\mu\nu}-K P_{\mu\nu}\right)\sqrt{-g^{00}}=0\,,
\end{align}
where the term proportional to arbitrary time-dependent function $\Gamma(t)$ induces dissipation, $P_{\mu\nu}$ is a spatial projector defined in \eqref{projector}, and $\Lambda(t)$, $c(t)$ and $M(t)$ are time-dependent functions. 

In parallel, Kaplanek, Mylova and Tolley showed in \cite{Kaplanek:2025moq} that gauge theories on the SK contour need to have double gauge invariance, acting both on retarded and advanced fields. This constraints had not been imposed in the general construction of \cite{Salcedo:2025ezu}. As we will discuss elsewhere, both observations in \cite{Christodoulidis:2025vxz} and \cite{Kaplanek:2025moq} can be used to refine the construction of \cite{Salcedo:2025ezu} and systematically write down a class of dissipative modifications of Einstein's gravity with the desired number of scalar degrees of freedom, which will be one for us. Here we study one simple example, which is different from \eqref{Perseas} but arises at the same order in derivatives acting on the metric. Using our algorithm, we show that this example has precisely one scalar degree of freedom. The presence of gauge identities and advanced gauge invariance will be crucial. The class of operators considered here actually displays classical equations that are proportional to equations that can be obtained from a specific closed EFT of dark energy. We leave this issue of open vs closed dynamics for future investigation and focus here on the counting the number of propagating degrees of freedom. \\




\subsubsection{A  minimal modification}

We begin by analysing the minimal modification of the Einstein equations,  written as follows:
\begin{equation}
	\E_{\mu\nu}\coloneqq
	G_{\mu\nu}
	+
	\frac{X}{f}\left( K_{\mu\nu}- K g_{\mu\nu}-n_\mu a_\nu - n_\nu a_\mu \right)
	-
	\frac{\dot X}{f} P_{\mu\nu}
	\label{modified Einstein EoMs}
\end{equation}
where $f=f(t)$ is a function of time, $P_{\mu\nu}$ is a projector symmetric tensor defined as
\begin{align}\label{projector}
	P_{\mu\nu}\coloneqq g_{\mu\nu}+n_\mu n_\nu \dot = g_{\mu\nu}+\frac{\delta^0_\mu \delta^0_\nu}{(-g^{00})} \,,
\end{align}
which projects onto spacelike hypersurfaces orthogonal to $n_\mu$, $P_{\mu\nu}n^\nu=0$ and $X$ is a function of the lapse $N=\frac{1}{\sqrt{-g^{00}}}$ defined as
\begin{align}
	X\coloneqq \frac{\dot f}{N}\,.
\end{align}
The last term contains the acceleration $a_\mu$, defined from the decomposition of the extrinsic curvature
\begin{align}
	K_{\mu\nu}=\nabla_\mu n_\nu+n_\mu a_\nu\,.
\end{align}
The derivation of \eqref{modified Einstein EoMs} will be presented elsewhere. Here we simply study it at face value. \\

As in the previous analysis, we perform an expansion up to linear order in metric perturbations. This allows us to extract the (modified) background equations at zeroth order, as well as the set of equations of motion governing the first-order perturbations. It is convenient to define
\begin{equation}
	\alpha\coloneqq \frac{\dot f}{f}\,,
	\qquad
	\beta\coloneqq \frac{\ddot f}{f}\,.
\end{equation}

 For clarity, we report here the relevant expansions of all terms appearing in \eqref{modified Einstein EoMs}, using the SVT-decomposition introduced above. 
\begin{align}
	&g^{00} = -1 + E, \qquad 
	N = 1 + \frac{E}{2}, \qquad 
	n_0 = -1 - \frac{E}{2}, \qquad 
	n_i = 0, \nonumber \\[4pt]
	&\frac{X}{f} = \frac{\dot f}{fN} = \alpha \left(1 - \frac{E}{2} \right), \qquad
	\frac{\dot X}{f} = \beta - \beta E - \frac{\alpha}{2} \dot E, \nonumber \\[6pt]
	&P_{00}=0,\qquad
	P_{0i}=a\,\partial_iF,\qquad
	P_{ij}=a^2\delta_{ij}+a^2\left(A\delta_{ij}+\partial_i\partial_jB\right), \nonumber \\[6pt]
	&a_0 = 0, \qquad  a_i = \frac{1}{2} \partial_i E, \label{useful expansions} \\[6pt]
	& K_{00} = 0, \qquad 
	  K_{0i} = a H\, \partial_i F, \qquad K = 3H+\frac{3}{2} (\dot A - H E) + \frac{k^2}{a} (F-\frac{a}{2}\dot B)\,, \nonumber \\[6pt]
     &K_{ij} = H a^2 \delta_{ij} + a^2 \left( \frac{1}{2} \dot A + H A - \frac{1}{2} H E \right)\delta_{ij} + \partial_i \partial_j (a \dot a  B + \frac{a^2}{2}\dot B -a F)\,, \nonumber
\end{align}
where we recall that $H= \dot a/a$ is the Hubble parameter. The background values of the Einstein tensor are given by
\begin{align}
	&\bar G_{00}=3H^2,\
	&\bar G_{ij}=-a^2(3H^2+2\dot H)\delta_{ij},.
\end{align}

The corresponding background equations then take the form
\begin{equation}
	\begin{cases}
		3H^2+3\alpha H=0\,,\\
		3H^2+2\dot H + 2\alpha H +\beta =0\,.
	\end{cases}
\end{equation}
 
 At the perturbative level, using the $\delta G_{\mu\nu}$ epression in \eqref{delta G_00}, \eqref{delta G_0i}, \eqref{delta G_ij} and the expansion in \eqref{useful expansions}, the scalar sector of the equations of motion can be written as follows:
  \begin{align}
 	&3H\dot A+\frac{k^2 A}{a^2}+2H k^2(a\dot B-F)
 	+\alpha\left(\frac32\dot A+k^2(a\dot B-F)\right)=0\,,
 	\\[4pt]
 	&\left(H+\frac{\alpha}{2}\right)E-\dot A=0\,,
 	\\[4pt]
 	&\left(H+\frac{\alpha}{2}\right)\dot E
 	-\left(3H+\alpha\right)\dot A
 	-\ddot A
 	-\frac{k^2(E+A)}{2a^2}+\nonumber\\
 	&\qquad \qquad k^2 \left(\frac{\ddot B}{2}+ \frac{3}{2}H \dot B -2 \frac{H}{a}F -\frac{1}{a}\dot F +\alpha(F-\frac{a}{2}\dot B) \right)
 	=0\,,
 	\\[4pt]
 	&-\frac12(E+A)+a^2\big(\frac{\ddot B}{2}+ \frac{3}{2}H \dot B -2 \frac{H}{a}F -\frac{1}{a}\dot F\big)- \frac{\alpha}{a}(F-\frac{a}{2}\dot B)=0\,.
 \end{align}
 
 Since both this system of equations and the expression for $K$ in \eqref{useful expansions} involve the fields $F$ and $B$ only through the combination $(F-\frac{a}{2}\dot B)$, we introduce a new field $\Sigma$, defined as
 \begin{align}
 	\Sigma \coloneqq (F-\frac{a}{2}\dot B)\,,
 \end{align}
 and recast the retarded equations of motion in terms of it as follows:
 \begin{align}
 	\E_1\coloneqq&3H\dot A+\frac{k^2 A}{a^2}+\frac{2H}{a}k^2\Sigma
 	+\alpha\left(\frac32\dot A+\frac{1}{a}k^2\Sigma\right)=0\,,
 	\\[4pt]
 	\E_2\coloneqq&\left(H+\frac{\alpha}{2}\right)E-\dot A=0\,,
 	\\[4pt]
 	\E_3\coloneqq&\left(H+\frac{\alpha}{2}\right)\dot E
 	-\left(3H+\alpha\right)\dot A
 	-\ddot A
 	-\frac{k^2(E+A)}{2a^2}
 	-\frac{1}{a}k^2\dot\Sigma
 	-\frac{2H+\alpha}{a}k^2\Sigma=0\,,
 	\\[4pt]
 	\E_4\coloneqq&-\frac12(E+A)-a\big(\dot\Sigma+(2H+\alpha)\Sigma\big)=0\,.
 \end{align}
  Note that, by using the field $\Sigma$ instead of $F$ and $B$, we have removed the gauge freedom associated with the transformation in \eqref{gauge transformation EFToI}, since $\Sigma$ is invariant under it. Of course, we did not need to do this, and we could have left the fields B and F as independent fields. Our algorithm would have then diagnosed the existence of this gauge redundancy. Studying the equations directly at the level of $\Sigma$ shortens the calculation and simplifies our presentation without changing any of the final results. 
  
  The retarded analysis produces four constraints and one gauge identity, so that 
 \begin{align}
    l_r&=4 \,, & g_r&=1 \,, & e_r&=2 \,.
\end{align}
 To perform the advanced analysis and verify the absence of gauge identities, we construct the MSR action by introducing an advanced field for each equation of motion:
 \begin{align}
 	\SSK = \int d^4x \sqrt{-g}\, \phi_a^i \E_i\,,
 \end{align}
 where $i = 1, \ldots, 4$. Note that, since there are four equations, we introduce four advanced fields, while there are only three independent retarded fields ($A, E, \Sigma$).
 
 The advanced analysis yields three constraints and no gauge identities, as expected.
 
 Finally, the total number of scalar degrees of freedom is one:
 \begin{align}
 	\#\text{d.o.f.} = N_r - \frac{1}{2}(l_r + g_a + e_a) = 3 - \frac{1}{2}(4) = 1\,.
 \end{align}


  \subsubsection{Additional operators}
  
 After constructing a consistent open modification, we now recover the effective field theory of inflation within this framework. To this end, we consider the following action
 \begin{align}
 	S^{(G)}[g]
 	=
 	\int d^4x\,\sqrt{-g}\,G(N,t),
 \end{align}
 where \(G(N)\) is a smooth function of the lapse \(N\), and \(G'(N)\coloneqq dG/dN\). This results in the following MSR action
 \begin{align}
 	\SSK^{(G)}
 	=
 	\frac12
 	\int d^4x\,\sqrt{-g}
 	\Bigl[
 	G(N,t)\,g_{\mu\nu}
 	-
 	N\,G'(N,t)\,n_\mu n_\nu
 	\Bigr]
 	a^{\mu\nu}.
 \end{align}
 The choice 
 \begin{align}
 	G(N)=1\,,
 \end{align}
reproduces the cosmological constant term, while the choice
\begin{align}
G(N)=-\frac{1}{N^2}\,,
\end{align}
reproduces the $c(t)$ term of the EFT of inflation. Since $G(N)$ does not contain derivatives of the metric, this contribution is unitary, as expected, since $S_{\text{universal}}$ is unitary.

We expand \(G(N)\) around the background value \(N=1\):
\begin{align}
    G(N) = G(1) + \frac{E}{2}\, G'(1)\,, 
    \qquad 
    G'(N) = G'(1) + \frac{E}{2}\, G''(1)\,.
    \label{G and G' expansion}
\end{align}

Introducing the short hand 
\begin{align}
   G_0\coloneqq G(1) \,, \qquad G_1 \coloneqq G'(1)\,,\qquad G_2 \coloneq G''(1) \,,
\end{align}
and using \eqref{G and G' expansion}, we obtain
\begin{equation}
G(N)=G_0+\frac{E}{2}G_1\,,
\qquad
N G'(N)\,n_0n_0
=
G_1+\frac{E}{2}(3G_1+G_2)\,.
\end{equation}
This leads to the following background contributions:
\begin{align}
    &(00): -(G_0+G_1)\,,\\
    &(ij): G_0a^2\delta_{ij}\,.
\end{align}
and to the following contributions at the perturbative level:
\begin{align}
(00)
&:
-\left(G_0+2G_1+\frac{G_2}{2}\right)E\,,
\\[3pt]
(0i)
&:G_0\,a\,\partial_iF\,,
\\[3pt]
(ij)
&:
a^2\left[
\left(G_0A+\frac{G_1}{2}E\right)\delta_{ij}
+G_0\,\partial_i\partial_jB
\right]\,.
\end{align}

We can now include this term in the modified action introduced in the previous section,
\begin{align}
    \SSK^\text{openGR}=\int\sqrt{-g}\ \E^\text{openGR}_{\mu\nu}a^{\mu\nu}\,,
\end{align}
where the $ \E^\text{openGR}_{\mu\nu}$ denotes the modification of the Einstein equations given in \eqref{modified Einstein EoMs}. This construction gives the full dissipative EFT of inflation.
The full MSR action is
\begin{align}
    \SSK^\text{openEFToI}=\SSK^\text{openGR}-\frac{1}{f}\SSK^\text{(G)}\,,
\end{align}
and the resulting equations of motion are given by
\begin{equation}
	\E_{\mu\nu}\coloneqq
	G_{\mu\nu}
	+
	\frac{X}{f}\left( K_{\mu\nu}- K g_{\mu\nu}-n_\mu a_\nu - n_\nu a_\mu \right)
	-
	\frac{\dot X}{f} P_{\mu\nu} -\frac{1}{2f} \Bigl[G(N,t)\,g_{\mu\nu} 	- N\,G'(N,t)\,n_\mu n_\nu \Bigr]\,.
	\label{modified Einstein EoMs plus universal}
\end{equation}
From this we can extract the modified background equations:
\begin{equation}
	\begin{cases}
		3H^2+3\alpha H+\frac{G_0+G_1}{2f}=0\,,\\
		3H^2+2\dot H + 2\alpha H +\beta+\frac{G_0}{2f} =0\,,
	\end{cases}
\end{equation}
and the scalar perturbation equations 
\begin{align}
\E_1\coloneqq
&
3H\dot A-\frac{\Delta A}{a^2}-\frac{2H}{a}\Delta\Sigma
+\frac{\dot f}{f}\left(\frac32\dot A-\frac{1}{a}\Delta\Sigma\right)
+\frac{1}{2f}\left(
G_0+2G_1+\frac{G_2}{2}
\right)E=0\,,
\\[6pt]
\E_2\coloneqq
&
\left(H+\frac{\dot f}{2f}\right)E-\dot A=0\,,
\\[6pt]
\E_3\coloneqq
&
\left(H+\frac{\dot f}{2f}\right)\dot E
-\left(3H+\frac{\dot f}{f}\right)\dot A
-\ddot A\ +
 \nonumber\\
&\qquad  +\frac{\Delta(E+A)}{2a^2}+\frac{1}{a}\Delta\dot\Sigma +\ \frac{1}{a}\left(2H+\frac{\dot f}{f}\right)\Delta\Sigma
-\frac{2G_0+G_1}{4f}\,E=0\,,
\\[6pt]
\E_4\coloneqq
&
-\frac12(E+A)-a\left(\dot\Sigma+\left(2H+\frac{\dot f}{f}\right)\Sigma\right)=0\,.
\end{align}
As anticipated, the analysis yields the same results as before. The retarded analysis gives four constraints and one gauge identity, so that
\begin{align}
    l_r &= 4 \,, & g_r &= 1 \,, & e_r &= 2 \,.
\end{align}
The gauge identity is unchanged and reads
\begin{equation}
\E_3
=
\dot{\E}_2
+
(3H+\alpha)\E_2
+
\frac{k^2}{a^2}\E_4\,.
\end{equation}

As in the previous case, the advanced analysis yields three constraints and no gauge identities.

Finally, the total number of scalar degrees of freedom is one:
\begin{align}
 	\#\text{d.o.f.} = N_r - \frac{1}{2}(l_r + g_a + e_a) = 3 - \frac{1}{2}(4) = 1\,.
\end{align}


\section{Conclusions and outlook}

In this paper we have developed an algorithm to count the number of degrees of freedom in open systems. This question is a necessary first step in constructing open effective field theories, but it is subtle whenever the equations of motion contain constraints, gauge redundancies, or dissipative terms. In ordinary conservative systems, the Hamiltonian Dirac--Bergmann algorithm, or its Lagrangian counterpart, provides a systematic answer (with caveats pointed out in \cite{Seiler:1995ne,Blagojevic:2020dyq,Hu:2022anq,Tomonari:2023wcs,Heisenberg:2025fxc}). By contrast, after integrating out environmental degrees of freedom, the effective dynamics is generically non-Hamiltonian and need not arise from the variation of an ordinary action. The standard counting formula therefore does not apply in general.

Our main result is a generalisation of the Lagrangian algorithm to \textit{classical, linear open systems}. We considered a set of linear equations of motion for retarded fields,
\[
        E_i^r = M_{iI}\phi^I_r = 0 ,
\]
where the number of equations need not equal the number of fields. The central step is to associate to this system a dual set of advanced equations, obtained equivalently by introducing the auxiliary Martin-Siggia-Rose functional
\[
        \SSK = \int \phi^i_a E^r_i
        = \int \phi^i_a M_{iI}\phi^I_r .
\]
The retarded equations determine the constraints on the retarded initial data, while the gauge redundancies of the retarded fields are controlled by the gauge identities of the advanced equations. This leads to the counting formula
\[
        \# {\rm d.o.f.}
        =
        N_r - {1\over 2}\left(l_r + g_a + e_a\right),
\]
where \(l_r\) is the number of retarded constraints, while \(g_a\) and \(e_a\) count the gauge identities and gauge transformations extracted from the advanced equations. This is the main difference with respect to the unitary formula: retarded gauge identities do not in general generate gauge redundancies of the retarded fields, but instead generate gauge redundancies of the advanced fields.

We have illustrated the algorithm in a variety of examples of increasing complexity. Simple coupled scalar systems make the distinction between retarded and advanced gauge identities transparent and show explicitly how the ordinary unitary counting can fail. Electromagnetism provides a useful benchmark, both before and after gauge fixing. In particular, gauge fixing at the level of the equations of motion naturally produces systems with unequal numbers of fields and equations, for which the retarded/advanced formulation is essential to recover the correct two propagating polarizations. We then applied the same logic to dissipative electromagnetic theories, including light in a medium, and found that the standard gauge redundancy of the retarded fields is correctly encoded in the advanced gauge identities. Finally, we considered gravitational effective theories relevant for cosmology, where the method provides a systematic way to identify the propagating scalar and tensor modes in the presence of constraints and gauge redundancies.

Besides counting degrees of freedom, the formalism also clarifies the role of advanced gauge symmetry. Retarded gauge identities imply gauge transformations of the auxiliary advanced fields. Although these do not remove physical retarded degrees of freedom, they have physical consequences. In particular, they impose consistency conditions on possible deterministic sources or stochastic noise terms. Thus, the algorithm not only identifies the number of propagating modes, but also constrains the allowed structure of stochastic extensions of the deterministic equations.

There are several natural directions for future work. 
\begin{itemize}
    \item It would be useful to study dissipative Proca theory in more detail. The massive vector provides a simple but non-trivial laboratory because the mass term removes the usual gauge redundancy while preserving constraint equations that eliminate the unphysical component of the vector field. A dissipative generalisation should propagate the correct number of degrees of freedom only for appropriate choices of the dissipative operators. Our algorithm can be used as a guiding principle to identify these consistent choices.
    \item It would be interesting to compare our construction with other recent approaches to counting degrees of freedom directly at the level of the equations of motion in \cite{Heisenberg:2025fxc}. In particular, methods based on the Cartan--Kuranishi analysis of differential equations are more general than the present formalism, since they can be applied to nonlinear partial differential equations. This generality comes at the price of increased calculational complexity. A detailed comparison in the common regime of linear ordinary differential equations would be valuable.
    \item The present work has been restricted to classical deterministic dynamics. A natural next step is to incorporate stochasticity and, ultimately, the full quantum Schwinger--Keldysh path integral. At the stochastic level, the advanced gauge symmetries identified here should impose non-trivial constraints on noise kernels and source correlations. At the quantum level, one would like to understand how the counting of degrees of freedom is reflected in the full doubled path integral, including higher powers of the advanced fields, fluctuation--dissipation relations, and the positivity constraints required for a consistent open quantum system. Interesting work in this direction for \textit{out-of-equilibrium} systems includes \cite{Calzetta:2004sh,Czajka:2014eha,Haehl:2016pec,Salcedo:2024nex,Kaplanek:2025moq,Kaplanek:2026kpp}. 
    \item Our original motivation was to understand dissipative generalisations of gravity. In such theories, one would like to know which operators can be added to the effective equations while preserving a desired spectrum of weakly coupled degrees of freedom. In cosmological applications, this is especially important for determining whether a theory contains a healthy, weakly coupled scalar mode in addition to the tensor polarizations of gravity \cite{Christodoulidis:2025vxz,Salcedo:2026cqb}. The formalism developed here provides a practical first step toward this goal: it gives an algorithmic way to diagnose constraints, gauge redundancies and consistency conditions directly from the equations of motion, without assuming the existence of an ordinary Hamiltonian or Lagrangian description.
\end{itemize}


\section*{Acknowledgements} We thank Thomas Colas and Eren Firat for helpful discussions and comments on the draft. E.~P.~is supported by STFC consolidated grants ST/T000694/1 and ST/X000664/1. E.~L.~acknowledges the hospitality of DAMTP.


\bibliographystyle{JHEP}
\bibliography{merged}

@article{Burgess:2015ajz,
    author = "Burgess, C. P. and Holman, R. and Tasinato, G.",
    title = "{Open EFTs, IR effects \& late-time resummations: systematic corrections in stochastic inflation}",
    eprint = "1512.00169",
    archivePrefix = "arXiv",
    primaryClass = "gr-qc",
    doi = "10.1007/JHEP01(2016)153",
    journal = "JHEP",
    volume = "01",
    pages = "153",
    year = "2016"
}

@article{Boyanovsky:2018fxl,
    author = "Boyanovsky, Daniel",
    title = "{Information loss in effective field theory: entanglement and thermal entropies}",
    eprint = "1801.06840",
    archivePrefix = "arXiv",
    primaryClass = "hep-th",
    doi = "10.1103/PhysRevD.97.065008",
    journal = "Phys. Rev. D",
    volume = "97",
    number = "6",
    pages = "065008",
    year = "2018"
}

@article{Cespedes:2025ple,
    author = "Cespedes, Sebastian and Qin, Zhehan and Wang, Dong-Gang",
    title = "{$\lambda \phi^4${\textasciitilde}as an Effective Theory in de Sitter}",
    eprint = "2510.25826",
    archivePrefix = "arXiv",
    primaryClass = "hep-th",
    reportNumber = "Imperial-TP-2025-SCC-3",
    month = "10",
    year = "2025"
}

@article{Christodoulidis:2025vxz,
    author = "Christodoulidis, Perseas and Gong, Jinn-Ouk",
    title = "{Gravitational open effective field theory of inflation}",
    eprint = "2512.21234",
    archivePrefix = "arXiv",
    primaryClass = "hep-th",
    month = "12",
    year = "2025"
}

@article{Salcedo:2025ezu,
    author = {Salcedo, Santiago Ag{\"u}{\'\i} and Colas, Thomas and Dufner, Lennard and Pajer, Enrico},
    title = "{An open system approach to gravity}",
    eprint = "2507.03103",
    archivePrefix = "arXiv",
    primaryClass = "hep-th",
    doi = "10.1007/JHEP02(2026)241",
    journal = "JHEP",
    volume = "02",
    pages = "241",
    year = "2026"
}

@article{Salcedo:2026sdn,
    author = {Salcedo, Santiago Ag{\"u}{\'\i} and Colas, Thomas and Suman, Petar and Zhang, Bowei and Fergusson, James and Shellard, E. P. S.},
    title = "{Primordial non-Gaussianity constraints on dissipative inflation}",
    eprint = "2603.13473",
    archivePrefix = "arXiv",
    primaryClass = "astro-ph.CO",
    month = "3",
    year = "2026"
}

@article{Heisenberg:2025fxc,
    author = "Heisenberg, Lavinia",
    title = "{Counting Degrees of Freedom: A Method Applicable from Scalars to f(Q) Gravity and Beyond}",
    eprint = "2509.18192",
    archivePrefix = "arXiv",
    primaryClass = "math-ph",
    month = "9",
    year = "2025"
}

@article{Green:2025hmo,
    author = "Green, Daniel and Gupta, Kshitij",
    title = "{Quantum Walks and Exact RG in de Sitter Space}",
    eprint = "2512.13842",
    archivePrefix = "arXiv",
    primaryClass = "hep-th",
    month = "12",
    year = "2025"
}

@article{Colas:2025ind,
    author = "Colas, Thomas and Qin, Zhehan and Tong, Xi",
    title = "{Open Effective Field Theory and the Physics of Cosmological Collider Signals}",
    eprint = "2512.07941",
    archivePrefix = "arXiv",
    primaryClass = "hep-th",
    month = "12",
    year = "2025"
}

@article{Mirbabayi:2024eml,
    author = "Mirbabayi, Mehrdad",
    title = "{Loosely coupled particles in warm inflation}",
    eprint = "2409.17927",
    archivePrefix = "arXiv",
    primaryClass = "astro-ph.CO",
    doi = "10.1088/1475-7516/2025/05/067",
    journal = "JCAP",
    volume = "05",
    pages = "067",
    year = "2025"
}

@article{Burgess:2014eoa,
    author = "Burgess, C. P. and Holman, R. and Tasinato, G. and Williams, M.",
    title = "{EFT Beyond the Horizon: Stochastic Inflation and How Primordial Quantum Fluctuations Go Classical}",
    eprint = "1408.5002",
    archivePrefix = "arXiv",
    primaryClass = "hep-th",
    reportNumber = "CERN-PH-TH-2014-142",
    doi = "10.1007/JHEP03(2015)090",
    journal = "JHEP",
    volume = "03",
    pages = "090",
    year = "2015"
}

@article{Boyanovsky:2015jen,
    author = "Boyanovsky, D.",
    title = "{Effective field theory during inflation. II. Stochastic dynamics and power spectrum suppression}",
    eprint = "1511.06649",
    archivePrefix = "arXiv",
    primaryClass = "astro-ph.CO",
    doi = "10.1103/PhysRevD.93.043501",
    journal = "Phys. Rev. D",
    volume = "93",
    pages = "043501",
    year = "2016"
}

@article{Boyanovsky:2015tba,
    author = "Boyanovsky, D.",
    title = "{Effective field theory during inflation: Reduced density matrix and its quantum master equation}",
    eprint = "1506.07395",
    archivePrefix = "arXiv",
    primaryClass = "astro-ph.CO",
    doi = "10.1103/PhysRevD.92.023527",
    journal = "Phys. Rev. D",
    volume = "92",
    number = "2",
    pages = "023527",
    year = "2015"
}

@article{Burgess:2022rdo,
    author = "Burgess, C. P. and Kaplanek, Greg",
    title = "{Gravity, Horizons and Open EFTs}",
    eprint = "2212.09157",
    archivePrefix = "arXiv",
    primaryClass = "hep-th",
    reportNumber = "CERN-TH-2022-175; Imperial/TP/2022/GK/03, CERN-TH-2022-175, Imperial/TP/2022/GK/03",
    month = "12",
    year = "2022"
}

@article{Hongo:2018ant,
    author = "Hongo, Masaru and Kim, Suro and Noumi, Toshifumi and Ota, Atsuhisa",
    title = "{Effective field theory of time-translational symmetry breaking in nonequilibrium open system}",
    eprint = "1805.06240",
    archivePrefix = "arXiv",
    primaryClass = "hep-th",
    reportNumber = "RIKEN-iTHEMS-Report-18, KOBE-COSMO-18-05, RIKEN-ITHEMS-REPORT-18",
    doi = "10.1007/JHEP02(2019)131",
    journal = "JHEP",
    volume = "02",
    pages = "131",
    year = "2019"
}

@article{Janssen:1976qag,
    author = "Janssen, Hans-Karl",
    title = "{On a Lagrangean for classical field dynamics and renormalization group calculations of dynamical critical properties}",
    doi = "10.1007/BF01316547",
    journal = "Z. Phys. B",
    volume = "23",
    number = "4",
    pages = "377--380",
    year = "1976"
}

@article{DeDominicis:1976,
  author  = {De Dominicis, C.},
  title   = {Techniques de renormalisation de la th{\'e}orie des champs
             et dynamique des ph{\'e}nom{\`e}nes critiques},
  journal = {Journal de Physique Colloques},
  volume  = {37},
  number  = {C1},
  pages   = {C1-247--C1-253},
  year    = {1976},
  doi     = {10.1051/jphyscol:1976138}
}

@article{PhysRevA.8.423,
  title = {Statistical Dynamics of Classical Systems},
  author = {Martin, P. C. and Siggia, E. D. and Rose, H. A.},
  journal = {Phys. Rev. A},
  volume = {8},
  issue = {1},
  pages = {423--437},
  numpages = {0},
  year = {1973},
  month = {Jul},
  publisher = {American Physical Society},
  doi = {10.1103/PhysRevA.8.423},
  url = {https://link.aps.org/doi/10.1103/PhysRevA.8.423}
}

@article{Liu:2018kfw,
    author = "Liu, Hong and Glorioso, Paolo",
    title = "{Lectures on non-equilibrium effective field theories and fluctuating hydrodynamics}",
    eprint = "1805.09331",
    archivePrefix = "arXiv",
    primaryClass = "hep-th",
    reportNumber = "MIT-CTP/5018; EFI-18-8, MIT-CTP-5018, EFI-18-8",
    doi = "10.22323/1.305.0008",
    journal = "PoS",
    volume = "TASI2017",
    pages = "008",
    year = "2018"
}

@article{Crossley:2015evo,
    author = "Crossley, Michael and Glorioso, Paolo and Liu, Hong",
    title = "{Effective field theory of dissipative fluids}",
    eprint = "1511.03646",
    archivePrefix = "arXiv",
    primaryClass = "hep-th",
    reportNumber = "MIT-CTP-4734",
    doi = "10.1007/JHEP09(2017)095",
    journal = "JHEP",
    volume = "09",
    pages = "095",
    year = "2017"
}

@article{LopezNacir:2011kk,
    author = "Lopez Nacir, Diana and Porto, Rafael A. and Senatore, Leonardo and Zaldarriaga, Matias",
    title = "{Dissipative effects in the Effective Field Theory of Inflation}",
    eprint = "1109.4192",
    archivePrefix = "arXiv",
    primaryClass = "hep-th",
    reportNumber = "SLAC-PUB-14995",
    doi = "10.1007/JHEP01(2012)075",
    journal = "JHEP",
    volume = "01",
    pages = "075",
    year = "2012"
}

@article{Sharifian:2023jem,
    author = "Sharifian, Mohammad and Zarei, Moslem and Abdi, Mehdi and Bartolo, Nicola and Matarrese, Sabino",
    title = "{Open quantum system approach to the gravitational decoherence of spin-1/2 particles}",
    eprint = "2309.07236",
    archivePrefix = "arXiv",
    primaryClass = "gr-qc",
    month = "9",
    year = "2023"
}

@article{Schwinger:1960qe,
    author = "Schwinger, Julian S.",
    title = "{Brownian motion of a quantum oscillator}",
    doi = "10.1063/1.1703727",
    journal = "J. Math. Phys.",
    volume = "2",
    pages = "407--432",
    year = "1961"
}

@article{Keldysh:1964ud,
    author = "Keldysh, L. V.",
    title = "{Diagram technique for nonequilibrium processes}",
    journal = "Zh. Eksp. Teor. Fiz.",
    volume = "47",
    pages = "1515--1527",
    year = "1964"
}

@article{Ota:2024mps,
    author = "Ota, Atsuhisa",
    title = "{Fluctuation-dissipation relation in cosmic microwave background}",
    eprint = "2402.07623",
    archivePrefix = "arXiv",
    primaryClass = "hep-th",
    month = "2",
    year = "2024"
}

@article{Glorioso:2017fpd,
    author = "Glorioso, Paolo and Crossley, Michael and Liu, Hong",
    title = "{Effective field theory of dissipative fluids (II): classical limit, dynamical KMS symmetry and entropy current}",
    eprint = "1701.07817",
    archivePrefix = "arXiv",
    primaryClass = "hep-th",
    reportNumber = "MIT-CTP-4860, EFI-17-2",
    doi = "10.1007/JHEP09(2017)096",
    journal = "JHEP",
    volume = "09",
    pages = "096",
    year = "2017"
}

@article{Hongo:2019qhi,
    author = "Hongo, Masaru and Kim, Suro and Noumi, Toshifumi and Ota, Atsuhisa",
    title = "{Effective Lagrangian for Nambu-Goldstone modes in nonequilibrium open systems}",
    eprint = "1907.08609",
    archivePrefix = "arXiv",
    primaryClass = "hep-th",
    reportNumber = "RIKEN-iTHEMS-Report-19, KOBE-COSMO-19-10",
    doi = "10.1103/PhysRevD.103.056020",
    journal = "Phys. Rev. D",
    volume = "103",
    number = "5",
    pages = "056020",
    year = "2021"
}

@article{Salcedo:2024smn,
    author = "Salcedo, Santiago Agui and Colas, Thomas and Pajer, Enrico",
    title = "{The Open Effective Field Theory of Inflation}",
    eprint = "2404.15416",
    archivePrefix = "arXiv",
    primaryClass = "hep-th",
    doi = "10.1007/JHEP10(2024)248",
    journal = "JHEP",
    volume = "10",
    pages = "248",
    year = "2024"
}

@inproceedings{Colas:2024lse,
    author = "Colas, Thomas",
    title = "{Open Effective Field Theories for cosmology}",
    booktitle = "{58th Rencontres de Moriond on Cosmology}",
    eprint = "2405.09639",
    archivePrefix = "arXiv",
    primaryClass = "astro-ph.CO",
    month = "5",
    year = "2024"
}

@article{Zarei:2021dpb,
    author = "Zarei, Moslem and Bartolo, Nicola and Bertacca, Daniele and Ricciardone, Angelo and Matarrese, Sabino",
    title = "{Non-Markovian open quantum system approach to the early Universe: Damping of gravitational waves by matter}",
    eprint = "2104.04836",
    archivePrefix = "arXiv",
    primaryClass = "astro-ph.CO",
    doi = "10.1103/PhysRevD.104.083508",
    journal = "Phys. Rev. D",
    volume = "104",
    number = "8",
    pages = "083508",
    year = "2021"
}

@phdthesis{Flauger:2009uta,
    author = "Flauger, Raphael Manfred",
    title = "{Constraining fundamental physics with cosmology}",
    school = "Texas U.",
    year = "2009"
}

@article{Vardhan:2024qdi,
    author = "Vardhan, Shreya and Grozdanov, Sa\v{s}o and Leutheusser, Samuel and Liu, Hong",
    title = "{Effective field theories of dissipative fluids with one-form symmetries}",
    eprint = "2408.12868",
    archivePrefix = "arXiv",
    primaryClass = "hep-th",
    reportNumber = "MIT-CTP/5752",
    month = "8",
    year = "2024"
}

@book{Weinberg:2008zzc,
    author = "Weinberg, Steven",
    title = "{Cosmology}",
    isbn = "978-0-19-852682-7",
    year = "2008", 
    publisher = "Oxford University Press"
}

@article{Calzetta:2004sh,
    author = "Calzetta, E. A.",
    editor = "Gunzig, E. and Mukhanov, Viatcheslav F. and Verdaguer, E.",
    title = "{The Two particle irreducible effective action in gauge theories}",
    eprint = "hep-ph/0402196",
    archivePrefix = "arXiv",
    doi = "10.1023/B:IJTP.0000048174.83795.3f",
    journal = "Int. J. Theor. Phys.",
    volume = "43",
    pages = "767--799",
    year = "2004"
}

@article{Czajka:2014eha,
    author = "Czajka, Alina and Mrowczynski, Stanislaw",
    title = "{Ghosts in Keldysh-Schwinger Formalism}",
    eprint = "1401.5773",
    archivePrefix = "arXiv",
    primaryClass = "hep-th",
    doi = "10.1103/PhysRevD.89.085035",
    journal = "Phys. Rev. D",
    volume = "89",
    number = "8",
    pages = "085035",
    year = "2014"
}

@article{Haehl:2016pec,
    author = "Haehl, Felix M. and Loganayagam, R. and Rangamani, Mukund",
    title = "{Schwinger-Keldysh formalism. Part I: BRST symmetries and superspace}",
    eprint = "1610.01940",
    archivePrefix = "arXiv",
    primaryClass = "hep-th",
    doi = "10.1007/JHEP06(2017)069",
    journal = "JHEP",
    volume = "06",
    pages = "069",
    year = "2017"
}

@article{Hongo:2024brb,
    author = "Hongo, Masaru and Sogabe, Noriyuki and Stephanov, Mikhail A. and Yee, Ho-Ung",
    title = "{Schwinger-Keldysh effective action for hydrodynamics with approximate symmetries}",
    eprint = "2411.08016",
    archivePrefix = "arXiv",
    primaryClass = "hep-th",
    reportNumber = "RIKEN-iTHEMS-Report-24",
    month = "11",
    year = "2024"
}

@article{Burgess:2024heo,
    author = "Burgess, C. P. and Colas, Thomas and Holman, R. and Kaplanek, Greg",
    title = "{Does decoherence violate decoupling?}",
    eprint = "2411.09000",
    archivePrefix = "arXiv",
    primaryClass = "hep-th",
    month = "11",
    year = "2024"
}

@article{Liu:2024tqe,
    author = "Liu, Yan and Sun, Ya-Wen and Wu, Xin-Meng",
    title = "{Holographic Schwinger-Keldysh effective field theories including a non-hydrodynamic mode}",
    eprint = "2411.16306",
    archivePrefix = "arXiv",
    primaryClass = "hep-th",
    month = "11",
    year = "2024"
}

@article{Kaplanek:2025moq,
    author = "Kaplanek, Greg and Mylova, Maria and Tolley, Andrew J.",
    title = "{Gauging Open EFTs from the top down}",
    eprint = "2512.17089",
    archivePrefix = "arXiv",
    primaryClass = "hep-th",
    reportNumber = "Imperial/TP/2025/AJT/1",
    month = "12",
    year = "2025"
}

@article{Salcedo:2024nex,
    author = "Salcedo, Santiago Agui and Colas, Thomas and Pajer, Enrico",
    title = "{An Open Effective Field Theory for light in a medium}",
    eprint = "2412.12299",
    archivePrefix = "arXiv",
    primaryClass = "hep-th",
    doi = "10.1007/JHEP03(2025)138",
    journal = "JHEP",
    volume = "03",
    pages = "138",
    year = "2025"
}

@article{Lau:2024mqm,
    author = "Lau, Pak Hang Chris and Nishii, Kanji and Noumi, Toshifumi",
    title = "{Gravitational EFT for dissipative open systems}",
    eprint = "2412.21136",
    archivePrefix = "arXiv",
    primaryClass = "hep-th",
    reportNumber = "KOBE-COSMO-24-06, UT-Komaba/24-11",
    month = "12",
    year = "2024"
}

@article{Lopez:2025arw,
    author = "Lopez, Francescopaolo and Bartolo, Nicola",
    title = "{Quantum signatures and decoherence during inflation from deep subhorizon perturbations}",
    eprint = "2503.23150",
    archivePrefix = "arXiv",
    primaryClass = "astro-ph.CO",
    month = "3",
    year = "2025"
}

@book{Rothe:2010dzf_dof,
    author = "Rothe, Heinz J. and Rothe, Klaus D.",
    title = "{Classical and quantum dynamics of constrained Hamiltonian systems}",
    publisher = "World Scientific",
    address = "Singapore",
    year = "2010"
}

@article{Dirac:1950pj,
    author = "Dirac, Paul A. M.",
    title = "{Generalized Hamiltonian dynamics}",
    doi = "10.4153/CJM-1950-012-1",
    journal = "Can. J. Math.",
    volume = "2",
    pages = "129--148",
    year = "1950"
}

@article{Tomonari:2023wcs,
    author = "Tomonari, Kyosuke and Bahamonde, Sebastian",
    title = "{Dirac{\textendash}Bergmann analysis and degrees of freedom of coincident f(Q)-gravity}",
    eprint = "2308.06469",
    archivePrefix = "arXiv",
    primaryClass = "gr-qc",
    doi = "10.1140/epjc/s10052-024-12677-x",
    journal = "Eur. Phys. J. C",
    volume = "84",
    number = "4",
    pages = "349",
    year = "2024",
    note = "[Erratum: Eur.Phys.J.C 84, 508 (2024)]"
}

@article{Hu:2022anq,
    author = "Hu, Kun and Katsuragawa, Taishi and Qiu, Taotao",
    title = "{ADM formulation and Hamiltonian analysis of f(Q) gravity}",
    eprint = "2204.12826",
    archivePrefix = "arXiv",
    primaryClass = "gr-qc",
    doi = "10.1103/PhysRevD.106.044025",
    journal = "Phys. Rev. D",
    volume = "106",
    number = "4",
    pages = "044025",
    year = "2022"
}

@article{Blagojevic:2020dyq,
    author = "Blagojevi{\'c}, Milutin and Nester, James M.",
    title = "{Local symmetries and physical degrees of freedom in $f(T)$ gravity: a Dirac Hamiltonian constraint analysis}",
    eprint = "2006.15303",
    archivePrefix = "arXiv",
    primaryClass = "gr-qc",
    doi = "10.1103/PhysRevD.102.064025",
    journal = "Phys. Rev. D",
    volume = "102",
    number = "6",
    pages = "064025",
    year = "2020"
}

@article{Anderson:1951ta,
    author = "Anderson, James L. and Bergmann, Peter G.",
    title = "{Constraints in covariant field theories}",
    doi = "10.1103/PhysRev.83.1018",
    journal = "Phys. Rev.",
    volume = "83",
    pages = "1018--1025",
    year = "1951"
}

@article{Seiler:1995ne,
    author = "Seiler, Werner M. and Tucker, Robin W.",
    title = "{Involution and constrained dynamics. 1: The Dirac approach}",
    eprint = "hep-th/9506017",
    archivePrefix = "arXiv",
    doi = "10.1088/0305-4470/28/15/022",
    journal = "J. Phys. A",
    volume = "28",
    pages = "4431--4452",
    year = "1995"
}

@book{Henneaux:1992ig,
    author = "Henneaux, M. and Teitelboim, C.",
    title = "{Quantization of gauge systems}",
    isbn = "978-0-691-03769-1",
    publisher = "Princeton University Press",
    year = "1992"
}

@article{Wipf:1993xg,
    author = "Wipf, Andreas W.",
    editor = "Ehlers, J. and Friedrich, H.",
    title = "{Hamilton's formalism for systems with constraints}",
    eprint = "hep-th/9312078",
    archivePrefix = "arXiv",
    reportNumber = "ETH-TH-93-48",
    doi = "10.1007/3-540-58339-4_14",
    journal = "Lect. Notes Phys.",
    volume = "434",
    pages = "22--58",
    year = "1994"
}

@article{Salcedo:2026cqb,
    author = {Salcedo, Santiago Ag{\"u}{\'\i} and Colas, Thomas and Dufner, Lennard and Pajer, Enrico},
    title = "{Phenomenology of an Open Effective Field Theory of Dark Energy}",
    eprint = "2603.12321",
    archivePrefix = "arXiv",
    primaryClass = "astro-ph.CO",
    month = "3",
    year = "2026"
}

@article{Kaplanek:2026kpp,
    author = "Kaplanek, Greg and Mylova, Maria and Tolley, Andrew J.",
    title = "{Schwinger-Keldysh Path Integral for Gauge theories}",
    eprint = "2604.26941",
    archivePrefix = "arXiv",
    primaryClass = "hep-th",
    month = "4",
    year = "2026"
}

\end{document}